\documentclass[12pt]{article}
\begin{document}

\title{More about Superparticle Sum Rules\\ 
in Grand Unified Theories}

\author{
Yoshiharu \textsc{Kawamura}\footnote{E-mail: haru@azusa.shinshu-u.ac.jp} 
and Teppei \textsc{Kinami}\footnote{E-mail: s06t303@shinshu-u.ac.jp} \\
{\it Department of Physics, Shinshu University, }\\
{\it Matsumoto 390-8621, Japan}
}

\date{
May 8, 2007}

\maketitle
\begin{abstract}
Sum rules among superparticle masses are derived under the assumption that models beyond the MSSM 
are four-dimensional supersymmetric grand unified theories or five-dimensional supersymmetric orbifold 
grand unified theories.
Sfermion sum rules are classified into four types and
those sum rules can be useful probes of the MSSM and beyond.
\end{abstract}

\section{Introduction}

The supersymmetric (SUSY) extension of the standard model has been attractive
as physics beyond the weak scale.\cite{N}
The naturalness problem can be partially solved by the introduction of supersymmetry.
The gauge coupling unification can be realized within the framework of the minimal
supersymmetric standard model (MSSM), under the assumption of $\lq$desert' between
the TeV scale and the unification scale.\cite{unif}
If the magnitude of soft SUSY breaking parameters is the weak or TeV scale,
superpartners will be discovered at the Large Hadron Collider (LHC) in the near future.
Then a next tast is to understand the structure of the MSSM (or an extension of the MSSM, e.g., the NMSSM) 
and disclose physics beyond the MSSM.
The MSSM contains many parameters, whose magnitudes can be determined by precision measurements.
Those experimental data will give a hint to explore the MSSM and beyond, including a mechanism of SUSY breaking 
and gauge symmetry breaking.
Hence it is important to investigate the particle spectrum and interactions beyond the standard model (SM). 
Much works using mass relations among scalar particles have been carried out 
based on such a motivation.\cite{FHK&N,M&R,KM&Y,P&P,K&T,Ch&H,K&M,A&P}

If the grand unification is indeed the case, then the theory can be described 
as a supersymmetric grand unified theory (SUSY GUT).\cite{SUSYGUT}
The theory is fascinating, but, in general, it suffers from problems related to Higgs multiplets,
e.g., triplet-doublet splitting problem and the problem of proton stability\cite{proton} 
in the minimal SUSY $SU(5)$ GUT.
Many interesting unified models have been constructed by an extention of gauge group and/or Higgs sector.
There is a possibility to solve the problem by the introduction of extra dimension.
The triplet-doublet splitting of Higgs multiplets, in fact, is elegantly realized based on SUSY $SU(5)$ GUT
in a five dimension.\cite{K,H&N}\footnote{
In four-dimensional heterotic string models, it has been pointed out that extra color triplets are projected out
by the Wilson line mechanism.\cite{IKN&Q}}
A variety of higher-dimensional SUSY GUTs on an orbifold have been proposed towards a construction of realistic model.
Recently, the possibility of complete family unification has been studied 
in five-dimensional SUSY $SU(N)$ orbifold GUTs.\cite{KK&O}
It is important to make distinctive predictions in order to specify models by using experimental data.

In this paper, we study the superparticle (sparticle) spectrum in the MSSM
and derive sum rules among sparticle masses
under the assumption that models beyond the MSSM are four-dimensional SUSY GUTs
or five-dimensional SUSY orbifold GUTs.
We classify sfermion sum rules and show that those sum rules can be useful probes of the MSSM and beyond.

The contents of this paper are as follows.
In section 2, general arguments regarding sparticle spectrum and sum rules are given based on the MSSM.
Sparticle sum rules are also examined under the assumption that scalar particles take a common 
soft SUSY breaking mass at some high energy scale as an example.
In section 3, specific sum rules among sparticle masses are derived from various SUSY GUTs.
Section 4 is devoted to conclusions and discussions.

\section{MSSM sparticle spectrum and sum rules}

\subsection{Basic assumptions and strategy}

First we list assumptions adopted in our analysis.\\
1. The theory beyond the SM is the MSSM.
Here the MSSM means the SUSY extention of the SM with the minimal particle contents,
without specifying the structure of soft SUSY breaking terms.
The superpartners and Higgs bosons have a mass whose magnitude is, at most, of order TeV scale.
We neglect the threshold correction at the TeV scale due to the mass difference among the MSSM particles.
Further the TeV scale is often identified with the weak scale ($M_{EW}$) for simplicity.\\
2. The MSSM holds from TeV scale to a high energy scale ($M_U$).
Above $M_U$, there is a new physics.
Possible candidates are SUSY GUTs and/or SUSY orbifold GUTs.
There is a big desert between $M_{EW}$ and $M_U$.
In a simple scenario, the grand unification occurs at $M_U = 2.1 \times 10^{16}$GeV.
We also consider a partial unification or the case that the grand unified gauge group is 
broken down to the SM one ($G_{SM}$) in several steps,
e.g., $G_{U} \to \cdots \to G_{n} \to G_{SM}$.
In such a case, the $M_U$ is the partial unification scale or the scale of 
gauge symmetry breaking at the final stage $G_{n} \to G_{SM}$.\\
3. The SUSY is broken in a SUSY breaking sector and the effect is mediated to our visible sector
as the appearance of soft SUSY breaking terms.
The pattern of soft SUSY breaking parameters reflects on the mechanism of SUSY breaking and/or the way of its mediation. 
We do not specify the mechanism of SUSY breaking.
In most cases, the gravity mediation is assumed.
We also assume that soft SUSY breaking terms respect the gauge invariance.
After the breakdown of gauge symmetry, there appear extra contributions to
soft SUSY breaking parameters, which do not respect the gauge symmetry any more.
For example, there are renormalization group (RG) effects, threshold corrections at the breaking scale, 
contributions from non-renormalizable interaction terms, 
$F$-term contributions (contributions from interactions in superpotential) and $D$-term contributions.
We assume that 1-loop RG effects and $D$-term contributions to soft SUSY breaking scalar masses 
dominate for the mass splitting in our analysis.
The $D$-term contributions, in general, originates from $D$-terms related to broken gauge symmetries 
when soft SUSY breaking parameters possess a non-universal structure and
the rank of gauge group lowers after the breakdown of gauge symmetry.\cite{Dterm,KM&Y}
In most cases, the magnitude of $D$-term condensation is, at most, of order TeV scale squared
and hence $D$-term contributions can induce sizable effects on sfermion spectrum.\\ 
4. The pattern of Yukawa couplings reflects flavor structure in unified theory.
We assume that a suitable pattern of Yukawa couplings is obtained in the low-energy effective theory.
We neglect effects of Yukawa couplings concerning to the first two generations and those of the off-diagonal ones
because they are small compared with the third generation ones.\\
5. The sufficient suppression of flavor-changing neutral currents (FCNC) processes 
requires the mass degeneracy for each squark and slepton species 
in the first two generations unless those masses are rather heavy 
or fermion and its superpartner mass matrices are aligned.
We assume that the generation-changing entries in the sfermion mass matrices are sufficiently small
in the basis where fermion mass matrices are diagonal.
At first, we derive sum rules without the requirement of mass degeneracy and after that we give a brief comment on
the case with the degenerate masses.\\
6. After some parameters are made real by the rephasing of fields, CP violation occurs if the rest are complex.
We assume that Yukawa couplings are dominant as a source of CP violation
and other parameters are real.

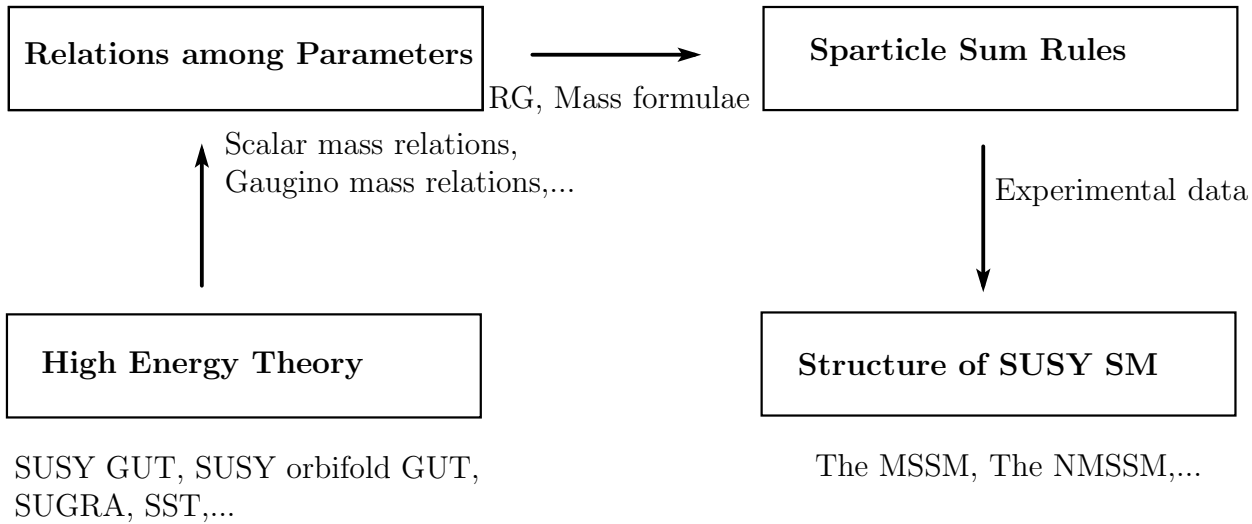
\begin{figure}
\caption{Outline of strategy}
\label{F1}
\unitlength 0.1in
\begin{picture}( 66.0000, 33.3000)( 10.3000,-34.1000)
%
\special{pn 13}%
\special{pa 400 380}%
\special{pa 2880 380}%
\special{pa 2880 920}%
\special{pa 400 920}%
\special{pa 400 380}%
\special{fp}%
\put(4.8000,-7.1000){\makebox(0,0)[lb]{\bf{Relations among Parameters}}}%
%
\special{pn 13}%
\special{pa 400 380}%
\special{pa 2880 380}%
\special{pa 2880 920}%
\special{pa 400 920}%
\special{pa 400 380}%
\special{fp}%
%
\special{pn 13}%
\special{pa 390 1990}%
\special{pa 2870 1990}%
\special{pa 2870 2530}%
\special{pa 390 2530}%
\special{pa 390 1990}%
\special{fp}%
%
\special{pn 13}%
\special{pa 4350 380}%
\special{pa 6830 380}%
\special{pa 6830 920}%
\special{pa 4350 920}%
\special{pa 4350 380}%
\special{fp}%
%
\special{pn 13}%
\special{pa 4340 1980}%
\special{pa 6820 1980}%
\special{pa 6820 2520}%
\special{pa 4340 2520}%
\special{pa 4340 1980}%
\special{fp}%
\put(5.7000,-23.3000){\makebox(0,0)[lb]{\bf{High Energy Theory}}}%
\put(45.9000,-6.9000){\makebox(0,0)[lb]{\bf{Sparticle Sum Rules}}}%
\put(45.3000,-23.1000){\makebox(0,0)[lb]{\bf{Structure of SUSY SM}}}%
%
\special{pn 20}%
\special{pa 1410 1840}%
\special{pa 1410 1130}%
\special{fp}%
\special{sh 1}%
\special{pa 1410 1130}%
\special{pa 1390 1198}%
\special{pa 1410 1184}%
\special{pa 1430 1198}%
\special{pa 1410 1130}%
\special{fp}%
\put(15.3000,-11.8000){\makebox(0,0)[lb]{Scalar mass relations,}}%
\put(15.3000,-13.8000){\makebox(0,0)[lb]{Gaugino mass relations,...}}%
\put(4.3000,-28.8000){\makebox(0,0)[lb]{SUSY GUT, SUSY orbifold GUT,}}%
\put(4.3000,-30.8000){\makebox(0,0)[lb]{SUGRA, SST,...}}%
\put(46.2000,-28.7000){\makebox(0,0)[lb]{The MSSM, The NMSSM,...}}%
%
\special{pn 20}%
\special{pa 3140 630}%
\special{pa 3990 630}%
\special{fp}%
\special{sh 1}%
\special{pa 3990 630}%
\special{pa 3924 610}%
\special{pa 3938 630}%
\special{pa 3924 650}%
\special{pa 3990 630}%
\special{fp}%
\put(29.1000,-9.4000){\makebox(0,0)[lb]{RG, Mass formulae}}%
%
\special{pn 20}%
\special{pa 5500 1110}%
\special{pa 5500 1840}%
\special{fp}%
\special{sh 1}%
\special{pa 5500 1840}%
\special{pa 5520 1774}%
\special{pa 5500 1788}%
\special{pa 5480 1774}%
\special{pa 5500 1840}%
\special{fp}%
\put(55.6000,-14.2000){\makebox(0,0)[lb]{Experimental data}}%
\end{picture}%
\end{figure}

Next we explain the outline of our strategy according to Figure \ref{F1}.
Let us construct a high energy theory with particular particle contents and (unified gauge) symmetries.
The theory, in general, contains free parameters including unknown quantities related to symmetry breakings, e.g., $D$-term contributions.
We can derive specific relations among soft SUSY breaking parameters at $M_U$
by eliminating free parameters.
Soft SUSY breaking parameters receive RG effects, and those values at $M_{EW}$ can be
calculated by using RG equations.
After the breakdown of electroweak symmetry, mass formulae of the physical masses are written, in terms of parameters in the SUSY SM.
Peculiar sum rules among sparticle masses at $M_{EW}$ are obtained by rewriting
specific relations at $M_U$ in terms of physical masses and parameters. 
In the near future, if superpartners and Higgs bosons were discoverd and those masses and interactions were precisely measured,
the presumed high energy theory can be tested by checking whether peculiar sum rules hold or not.

\subsection{MSSM sparticle spectrum}

We review the particle spectrum in the MSSM.\cite{B&T}
The Higgs sector consists of two Higgs doublets $H_{1}(=H_{d})$ and $H_{2}(=H_{u})$.
The stationary conditions of Higgs potential are given by
\begin{eqnarray}
M_{Z}^{2} = -\frac{m_{1}^{2}-m_{2}^{2}}{\cos 2\beta}-\left(m_{1}^{2}+m_{2}^{2}\right) ,~~
m_{1}^{2}+m_{2}^{2} = -\frac{2\mu B}{\sin 2\beta} ,
\label{B}
\end{eqnarray}
where $m_1$ and $m_2$ are defined by $m_1^2 \equiv \mu^2 + m_{H_1}^2$ and $m_2 \equiv \mu^2 + m_{H_2}^2$, 
and $B$ is the Higgs mass mixing parameter.
Here $m_{H_1}$ and  $m_{H_2}$ are soft SUSY breaking masses of Higgs doublets. 
The angle $\beta$ is defined by the ratio of the vacuum expectation values (VEVs) of neutral components of
Higgs bosons such that $\tan \beta \equiv {v_{2}}/{v_{1}}$. 
After the breakdown of electroweak symmetry,
one physical charged Higgs boson ($H^{\pm}$) and three kinds of neutral Higgs bosons ($H^{0}$, $h^{0}$, $A^{0}$) appear. 
The mass-squared of pseudoscalar Higgs boson ($A^0$) is given by 
\begin{eqnarray}
M_{A}^{2} = m_{1}^{2}+m_{2}^{2} = 2 \mu^2 + m_{H_1}^{2}+m_{H_2}^{2} .
\label{MA}
\end{eqnarray}
Using Eqs.(\ref{B}) and (\ref{MA}), we derive relations\footnote{
In the case that $\tan \beta$ is large, i.e.,
$\cos 2\beta = \frac{1-\tan^{2}\beta}{1+\tan^{2}\beta} \approx -1$ and
$\sin 2\beta = \frac{2\tan\beta}{1+\tan^{2}\beta} \approx \frac{2}{\tan\beta}$,
we have the approximation formulae
$B \approx - \frac{M_A^2}{\mu \tan\beta}$, 
$m_{1}^{2} \approx M_{A}^{2}+\frac{1}{2}M_{Z}^{2}$ and
$m_{2}^{2} \approx -\frac{1}{2}M_{Z}^{2}$.}
\begin{eqnarray}
&~& B = -\frac{\sin 2\beta}{2\mu} M_A^2 ,~
\label{B2}\\
&~& m_{1}^{2} = \frac{1}{2}\left(M_{A}^{2}- \left(M_{A}^{2}+M_{Z}^{2}\right)\cos 2\beta\right) ,
\label{m1}\\
&~& m_{2}^{2} = \frac{1}{2}\left(M_{A}^{2}+\left(M_{A}^{2}+M_{Z}^{2}\right)\cos 2\beta\right) .
\label{m2}
\end{eqnarray}
Using the above relations (\ref{B2})-(\ref{m2}),  values of $B$, $m_1$ and $m_2$ can be fixed.
There exist sum rules among Higgs masses such that
\begin{eqnarray}
\hspace{-10mm} &~& M_{H^{\pm}}^{2} = M_{A}^{2} + M_{W}^{2} ,~
\label{SR-Higgs1}\\
\hspace{-10mm} &~& M_{h}^{2}+M_{H}^{2} = M_{A}^{2}+M_{Z}^{2} + \delta_h ,~
\label{SR-Higgs2}\\
\hspace{-10mm} &~& M_{h}^2 M_{H}^{2} = M_{A}^{2}M_{Z}^{2}\cos^{2}2\beta 
 + \frac{\delta_h}{2}\left(M_{A}^{2}+M_{Z}^{2} + (M_{A}^{2}-M_{Z}^{2})\cos 2\beta\right) ,
\label{SR-Higgs3}
\end{eqnarray}
where $M_{H^{\pm}}$, $M_{H}$ and $M_{h}$ are masses of $H^{\pm}$, $H^0$ and $h^0$,
and 1-loop corrections to the scalar potential are considered.
The $\delta_h$ represents the radiative corrections,
\begin{eqnarray}
\delta_h = \frac{3g^2 m_t^4}{16\pi^2 M_W^2 \sin^2\beta}
 \ln\left(\left(1+\frac{M_{\tilde{t}_L}^2}{m_t^2}\right)\left(1+\frac{M_{\tilde{t}_R}^2}{m_t^2}\right)\right) .
\label{delta-h}
\end{eqnarray}
Here $m_t$ and $M_{\tilde{t}_{L(R)}}^2$ stand for top quark mass and scalar top quark masses, respectively.
For simplicity, we ignore the intra-generational mixing such as $\tilde{t}_L$-$\tilde{t}_R$ mixing
and contributions from bottom Yukawa coupling.
Hence the Higgs sector in the MSSM can be tested by the above mass formulae and sum rules.

Next we consider superpartners of gauge and Higgs bosons.
The gaugino masses relating to the $SU(3)_C \times SU(2)_L \times U(1)_Y$ gauge group 
are denoted $M_i$ $(i = 3, 2, 1)$ with a GUT normalization for weak hypercharge 
and the SUSY Higgs mass (or higgsino mass) parameter called $\lq$$\mu$ parameter' is denoted $\mu$.
The mass eigenstates in gaugino and higgsino sector consist of gluino ($\tilde{g}$)
whose mass is given by $M_{\tilde{g}} = M_{3}$ at tree level,\footnote{
It is known that the gluino mass receives relatively large radiative corrections such that
$M_{\tilde{g}} = M_3(Q)(1 + \frac{\alpha_3}{4\pi}(15 + 6 \ln(Q/M_3) + \sum A_q)$ where
$Q$ is the renormalization scale and  
$\sum A_q$ is a sum of contributions from quark and squark loops.\cite{M&V}} and 
two kinds of mixture of the gauginos and higgsinos, i.e., charginos ($\tilde{\chi}_{i}^{\pm}; i=1,2$) 
and neutralinos ($\tilde{\chi}_{i}^{0}; i=1,2,3,4$). 
The mass-squareds of charginos are given by
\begin{eqnarray}
&~& M_{\tilde{\chi}_{1}^{\pm}}^2 = \frac{1}{2}\left(M_{2}^{2}+\mu^{2} +2M_{W}^{2}-\Delta_{C}\right) ,~
\label{Mchi1}\\
&~& M_{\tilde{\chi}_{2}^{\pm}}^2 = \frac{1}{2}\left(M_{2}^{2}+\mu^{2} +2M_{W}^{2}+\Delta_{C}\right) ,
\label{Mchi2}
\end{eqnarray}
where $\Delta_{C}$ is defined as 
\begin{eqnarray}
\Delta_{C} \equiv \sqrt{\left(M_{2}^{2} + \mu^{2} +2M_{W}^{2}\right)^{2} -4\left(M_{2} \mu -M_{W}^{2}\sin 2\beta\right)^{2}} .
\label{DeltaC}
\end{eqnarray}
{}From Eqs.(\ref{Mchi1}) and (\ref{Mchi2}), the following relations are derived as
\begin{eqnarray}
&~& M_{\tilde{\chi}_{1}^{\pm}}^{2} + M_{\tilde{\chi}_{2}^{\pm}}^{2} = M_{2}^{2} + \mu^{2} + 2M_{W}^{2} ,
\label{chargino1}\\
&~& M_{\tilde{\chi}_{1}^{\pm}}^{2} M_{\tilde{\chi}_{2}^{\pm}}^{2} = \left(M_{2} \mu -M_{W}^{2} \sin 2\beta\right)^{2} .
\label{chargino2}
\end{eqnarray}
In the same way, relations among neutralino masses are given by
\begin{eqnarray}
\hspace{-10mm} &~& M_{\tilde{\chi}_{1}^{0}}^{2} + M_{\tilde{\chi}_{2}^{0}}^{2} + M_{\tilde{\chi}_{3}^{0}}^{2} + M_{\tilde{\chi}_{4}^{0}}^{2}
= M_{1}^{2} + M_{2}^{2} +2\mu^{2} +2M_{Z}^{2} ,
\label{neutralino1}\\
\hspace{-10mm} &~& M_{\tilde{\chi}_{1}^{0}}^{2} M_{\tilde{\chi}_{2}^{0}}^{2} + M_{\tilde{\chi}_{1}^{0}}^{2} M_{\tilde{\chi}_{3}^{0}}^{2} 
+ M_{\tilde{\chi}_{1}^{0}}^{2} M_{\tilde{\chi}_{4}^{0}}^{2} + M_{\tilde{\chi}_{2}^{0}}^{2} M_{\tilde{\chi}_{3}^{0}}^{2} 
+ M_{\tilde{\chi}_{2}^{0}}^{2} M_{\tilde{\chi}_{4}^{0}}^{2} + M_{\tilde{\chi}_{3}^{0}}^{2} M_{\tilde{\chi}_{4}^{0}}^{2} 
\nonumber \\
\hspace{-10mm} &~& ~~~~ = \left(M_1 M_2 - \mu^2 - M_Z^2\right)^2 
+ 2\left(M_{1} + M_{2}\right) \left(\mu^2(M_{1} + M_{2})  \right.
\nonumber \\
\hspace{-10mm} &~& ~~~~~~~~~~~~~~~~ \left. + M_W^2 (M_{1} + M_{2}\tan^2\theta_W) - \mu M_{Z}^{2}\sin2\beta\right)
\nonumber \\
\hspace{-10mm} &~& ~~~~~~~~~~~ - 2 \mu^2 M_1 M_2 + 2 \mu M_W^2 (M_{1} + M_{2}\tan^2\theta_W) \sin 2\beta  ,
\label{neutralino2}\\
\hspace{-10mm} &~& M_{\tilde{\chi}_{1}^{0}}^{2} M_{\tilde{\chi}_{2}^{0}}^{2} M_{\tilde{\chi}_{3}^{0}}^{2} 
+ M_{\tilde{\chi}_{1}^{0}}^{2} M_{\tilde{\chi}_{2}^{0}}^{2} M_{\tilde{\chi}_{4}^{0}}^{2} 
+ M_{\tilde{\chi}_{1}^{0}}^{2} M_{\tilde{\chi}_{3}^{0}}^{2} M_{\tilde{\chi}_{4}^{0}}^{2} 
+ M_{\tilde{\chi}_{2}^{0}}^{2} M_{\tilde{\chi}_{3}^{0}}^{2} M_{\tilde{\chi}_{4}^{0}}^{2} 
\nonumber \\
\hspace{-10mm} &~& ~~~~ = \left(\mu^2(M_{1} + M_{2}) + M_W^2 (M_{1} + M_{2}\tan^2\theta_W) - \mu M_Z^2 \sin2\beta\right)^2
\nonumber \\
\hspace{-10mm} &~& + 2\left(\mu^2 M_1 M_2 - \mu M_W^2 (M_{1} + M_{2}\tan^2\theta_W) \sin 2\beta\right)
  \left(M_{1}  M_{2} - \mu^2 - M_Z^2\right) ,
\label{neutralino3}\\
\hspace{-10mm} &~& M_{\tilde{\chi}_{1}^{0}}^{2}  M_{\tilde{\chi}_{2}^{0}}^{2} 
 M_{\tilde{\chi}_{3}^{0}}^{2} M_{\tilde{\chi}_{4}^{0}}^{2}
 = \left(-\mu^{2}M_{1}M_{2} +\mu M_{W}^{2}\left(M_{1}+M_{2}\tan^{2}\theta_{W}\right)\sin 2\beta\right)^{2} .
\label{neutralino4}
\end{eqnarray}
Using the above relations, values of $M_{2}^2$, $M_1^2$, $\mu^{2}$ and $\tan \beta$ can be determined. 
Hence the gaugino and higgsino sector in the MSSM can be tested by those sum rules.

Finally we consider the matter sector.
After the breakdown of electroweak symmetry, two kinds of contributions are added to sfermion masses, i.e.,
fermion masses ($m_f$) and the $D$-term contribution ($D_{W}(\tilde{f})$) relating to 
the generator of the broken symmetry ($SU(2)_{L} \times U(1)_{Y})/U(1)_{EM}$.
The diagonal elements ($M_{\tilde{f}}^2$) of sfermion mass-squared matrices are written as
\begin{eqnarray}
M_{\tilde{f}}^{2} = m_{\tilde{F}}^{2} + m_{f}^{2} + D_{W}(\tilde{f}) ,
\label{Mf}
\end{eqnarray}
where $f$ labels fermion species such that
\begin{eqnarray}
f = \left\{
\begin{array}{ccccccc}
{u}_{L},& {d}_{L},& {u}_{R},& {d}_{R},& {\nu}_{eL},& {e}_{L},& {e}_{R},\\
{c}_{L},& {s}_{L},& {c}_{R},& {s}_{R},& {\nu}_{\mu L},& {\mu}_{L},& {\mu}_{R},\\
{t}_{L},& {b}_{L},& {t}_{R},& {b}_{R},& {\nu}_{\tau L},& {\tau}_{L},& {\tau}_{R} ,
\end{array}
\right. 
\label{f}
\end{eqnarray}
and $\tilde{f}$ means the scalar partner of $f$.
The $m_{\tilde{F}}^{2}$ is the soft SUSY breaking mass of scalar particle $\tilde{F}$.
The $\tilde{F}$ represents a multiplet of $G_{SM}$, which contains
the scalar partner of $f$, and are given by,
\begin{eqnarray}
\tilde{F} = \left\{
\begin{array}{ccccc}
\tilde{q}_{1},& \tilde{u}_{R}^*,&\tilde{d}_{R}^*,&\tilde{l}_{1},&\tilde{e}_{R}^*,\\
\tilde{q}_{2},& \tilde{c}_{R}^*,&\tilde{s}_{R}^*,&\tilde{l}_{2},&\tilde{\mu}_{R}^*,\\
\tilde{q}_{3},& \tilde{t}_{R}^*,&\tilde{b}_{R}^*,&\tilde{l}_{3},&\tilde{\tau}_{R}^* ,
\end{array}
\right. 
\label{F}
\end{eqnarray}
where $\tilde{q}_1$ means the first generation scalar quark (squark) doublet, 
$\tilde{u}^*_R$ up squark singlet, $\tilde{d}^*_R$ down squark singlet,
$\tilde{l}_1$ the first generation scalar lepton (slepton) doublet, $\tilde{e}^*_R$ selectron singlet and so on.
The astrisk means its complex conjugate.
The $D_{W}(\tilde{f})$ are given by
\begin{eqnarray}
D_{W}(\tilde{f}) &=& \left(T_{L}^{3}(\tilde{f})-Q(\tilde{f})\sin^{2}\theta_{W}\right) M_{Z}^{2}\cos 2\beta 
\nonumber \\
&=& \left(\left(T_{L}^{3}-Q(\tilde{f})\right)M_{Z}^{2}+Q(\tilde{f})M_{W}^{2}\right)\cos 2\beta ~~ (f = u_L, \cdots \tau_L) ,
\label{DfL}\\
D_{W}(\tilde{f}) &=&  Q(\tilde{f})\sin^{2}\theta_{W} M_{Z}^{2}\cos 2\beta 
\nonumber \\
&=& Q(\tilde{f})\left(M_{Z}^{2} - M_{W}^{2}\right)\cos 2\beta ~~~~~~~~~~~~~~~~~~ (f = u_R, \cdots \tau_R) .
\label{DfR}
\end{eqnarray}
The off-diagonal elements of sfermion mass-squared matrices are proportional to the corresponding fermion mass.
For  the first two generations, the diagonal ones $M_{\tilde{f}}^{2}$ are regarded as $\lq$physical masses' 
which are eigenvalues of mass-squared matrices because the off-diagonal ones are negligibly small.
Using the mass formula (\ref{Mf}), values of ${m}_{\tilde{F}}^{2}$ can be determined for the first two generations.
For the third generation, mass-squared matrices are given by
\begin{eqnarray}
\hspace{-13mm} &~& \left(
\begin{array}{cc}
 m_{\tilde{t}_L}^{2} + m_{t}^{2} + D_{W}(\tilde{t}_L) & -m_t(A_t + \mu \cot\beta) \\
 -m_t(A_t + \mu \cot\beta) & m_{\tilde{t}_R}^{2} + m_{t}^{2} + D_{W}(\tilde{t}_R) 
\end{array}
\right) ~~ (\mbox{for top squarks}),
\label{mass-stop}\\
\hspace{-13mm} &~& \left(
\begin{array}{cc}
 m_{\tilde{b}_L}^{2} + m_{b}^{2} + D_{W}(\tilde{b}_L) & -m_b(A_b + \mu \tan\beta) \\
 -m_b(A_b + \mu \tan\beta) & m_{\tilde{b}_R}^{2} + m_{b}^{2} + D_{W}(\tilde{b}_R) 
\end{array}
\right)  (\mbox{for bottom squarks}),
\label{mass-sbottom}\\
\hspace{-13mm} &~& \left(
\begin{array}{cc}
 m_{\tilde{\tau}_L}^{2} + m_{\tau}^{2} + D_{W}(\tilde{\tau}_L) & -m_{\tau}(A_{\tau} + \mu \tan\beta) \\
 -m_{\tau}(A_{\tau} + \mu \tan\beta) & m_{\tilde{\tau}_R}^{2} + m_{\tau}^{2} + D_{W}(\tilde{\tau}_R) 
\end{array}
\right) ~~ (\mbox{for tau sleptons}),
\label{mass-stau}
\end{eqnarray}
where $A_t$, $A_b$ and $A_{\tau}$ are trilinear coupling parameters called $\lq$$A$ parameter'.
By diagonalized the above mass-squared matrices, we obtain mass eigenstates whose masses are physical ones,
($M_{\tilde{t}_1}$, $M_{\tilde{t}_2}$) for top squarks, ($M_{\tilde{b}_1}$, $M_{\tilde{b}_2}$) for bottom squarks
and ($M_{\tilde{\tau}_1}$, $M_{\tilde{\tau}_2}$) for tau sleptons.
By using the feature of trace, we have the relations,
\begin{eqnarray}
&~& M_{\tilde{t}_1}^2 + M_{\tilde{t}_2}^2 = M_{\tilde{t}_L}^2 + M_{\tilde{t}_R}^2 , ~~
M_{\tilde{b}_1}^2 + M_{\tilde{b}_2}^2 = M_{\tilde{b}_L}^2 + M_{\tilde{b}_R}^2 ,
\nonumber \\
&~& M_{\tilde{\tau}_1}^2 + M_{\tilde{\tau}_2}^2 = M_{\tilde{\tau}_L}^2 + M_{\tilde{\tau}_R}^2 .
\label{trace}
\end{eqnarray}
By diagonalizing the mass squared matrices, we have the relations,
\begin{eqnarray}
&~& \left(M_{\tilde{t}_1}^2 - M_{\tilde{t}_2}^2\right)^2 = 
 \left(M_{\tilde{t}_L}^2 - M_{\tilde{t}_R}^2\right)^2 + 4 m_t^2 \left(A_t + \mu \cot\beta\right)^2 , ~~
\label{t-}\\
&~& \left(M_{\tilde{b}_1}^2 - M_{\tilde{b}_2}^2\right)^2 = 
 \left(M_{\tilde{b}_L}^2 - M_{\tilde{b}_R}^2\right)^2 + 4 m_b^2 \left(A_b + \mu \tan\beta\right)^2 , ~~
\label{b-}\\
&~& \left(M_{\tilde{\tau}_1}^2 - M_{\tilde{\tau}_2}^2\right)^2 = 
 \left(M_{\tilde{\tau}_L}^2 - M_{\tilde{\tau}_R}^2\right)^2 + 4 m_{\tau}^2 \left(A_{\tau} + \mu \tan\beta\right)^2 , ~~
\label{t-}
\end{eqnarray}
If $A$ parameters were measured precisely,
${m}_{\tilde{F}}^{2}$s (and ${M}_{\tilde{f}}^{2}$s) in the third generation can be fixed 
by using the mass-squared matrices (\ref{mass-stop}) - (\ref{mass-stau}).
{}From the fact that left-handed fermions (and its superpartners) form $SU(2)_L$ doublets, e.g., $q_1 = (u_L, d_L)$ 
(and $\tilde{q}_1 = (\tilde{u}_L, \tilde{d}_L)$),
we obtain following sum rules among $SU(2)_L$ doublet sfermions:
\begin{eqnarray}
&~& M_{\tilde{u}_L}^2 - M_{\tilde{d}_L}^2 = m_u^2 - m_d^2 + M_W^2 \cos2\beta \simeq  M_W^2 \cos2\beta ,
\label{u-d} \\
&~& M_{\tilde{\nu}_{eL}}^2 - M_{\tilde{e}_L}^2 = m_{\nu_{eL}}^2 - m_e^2 + M_W^2 \cos2\beta \simeq  M_W^2 \cos2\beta ,
\label{nu-e} \\
&~& M_{\tilde{c}_L}^2 - M_{\tilde{s}_L}^2 = m_c^2 - m_s^2 + M_W^2 \cos2\beta \simeq  M_W^2 \cos2\beta ,
\label{c-s} \\
&~& M_{\tilde{\nu}_{\mu L}}^2 - M_{\tilde{\mu}_L}^2 = m_{\nu_{\mu L}}^2 - m_\mu^2 + M_W^2 \cos2\beta \simeq  M_W^2 \cos2\beta ,
\label{nu-mu} \\
&~& M_{\tilde{t}_L}^2 - M_{\tilde{b}_L}^2 = m_t^2 - m_b^2 + M_W^2 \cos2\beta \simeq  m_t^2 + M_W^2 \cos2\beta ,
\label{t-b} \\
&~& M_{\tilde{\nu}_{\tau L}}^2 - M_{\tilde{\tau}_L}^2 = m_{\nu_{\tau L}}^2 - m_\tau^2 + M_W^2 \cos2\beta \simeq  M_W^2 \cos2\beta ,
\label{nu-tau} 
\end{eqnarray}
where we neglect fermion masses except for the top quark mass in the final expressions.
The above sum rules (\ref{u-d}) - (\ref{nu-tau}) are irrelevant to the structure of models beyond the MSSM, and hence
the sfermion sector (and the breakdown of electroweak symmetry) in the MSSM can be tested by using them.
We refer these sum rules (\ref{u-d}) - (\ref{nu-tau}) as the type I sfermion sum rules.

\subsection{Renormalization group evolution of parameters}

We review RG evolution of coupling constants and mass parameters at 1-loop level.\cite{B&T}
We use conventional RG equations (RGEs) of soft SUSY breaking parameters 
in the case that the dynamics in the hidden sector do not give sizable effects.
\footnote{In Ref. \cite{CR&S}, Cohen et al. have pointed out that the conventional 
RGEs are modified by the hidden dynamics until auxiliary fields in the hidden sector freeze into their VEV 
unless interactions in the hidden sector are suppressed enough, and derived mass relations among scalar fields.}

RGEs regarding gauge couplings $g_i$ are given by
\begin{eqnarray}
\frac{{\rm d}\ }{{\rm d}t}\alpha_{i} \equiv - 2 \pi Q \frac{\partial}{\partial Q}\alpha_i =  -b_{i}\alpha_{i}^{2} ,~ 
\label{g-RGE}
\end{eqnarray}
where $\alpha_i \equiv g_i^2/(4\pi)$, $Q$ is the renormalization scale, $b_{i}$s are coefficients of beta functions at 1-loop level, i.e.,
$b_{3} = -3, b_{2} = 1, b_{1} = 33/5$ in the MSSM. 
By solving Eq.(\ref{g-RGE}), we have evolution of gauge couplings such that
\begin{eqnarray}
\alpha_i(M_U) = \frac{\alpha_{i}(Q)}{1-\frac{b_{i}}{2\pi}\alpha_{i}(Q)\ln({M_U}/{Q})}.
\label{g-sol}
\end{eqnarray}

The RGEs of gaugino masses are given by
\begin{eqnarray}
\frac{{\rm d}\ }{{\rm d}t}\left(\frac{M_{i}}{\alpha_{i}}\right) = 0
\label{Mi-RGE}
\end{eqnarray}
and Eq.(\ref{Mi-RGE}) is easily solved as
\begin{eqnarray}
&~& M_{i}(Q) = \frac{\alpha_{i}(Q)}{\alpha_i(M_U)}M_{i}(M_U)
= \left(1-\frac{b_{i}}{2\pi}\alpha_{i}(Q)\ln\frac{M_U}{Q}\right)M_{i}(M_U).
\label{Mi-sol}
\end{eqnarray}

In the first two generations, RGEs concerning soft SUSY breaking sfermion masses are given by 
\begin{eqnarray}
&~& \frac{{\rm d}\ }{{\rm d}t}m_{\tilde{F}}^{2} = 4 \sum_{i=1}^{3}C_2^{(i)}(\tilde{F}) \alpha_{i}M_{i}^{2}
 -\frac{3}{5}Y(\tilde{F})\alpha_{1}S ,
\label{mF-RGE}\\
&~& \frac{{\rm d}\ }{{\rm d}t}S = -b_{1}\alpha_{1}S ,~ S \equiv \sum_{\tilde{F}}Y(\tilde{F})n_{\tilde{F}}m_{\tilde{F}}^{2} ,
\label{S-RGE}
\end{eqnarray}
where $C_2^{(i)}(\tilde{F})$ and $Y(\tilde{F})$ represent the eigenvalues of second Casimir operator 
(e.g., $C_2^{(1)}({\tilde{q}_1})=4/3$, $C_2^{(2)}({\tilde{q}_1})=3/4$ and $C_2^{(1)}({\tilde{q}_1})=1/60$)
and hypercharge for $\tilde{F}$, respectively. 
In Eq.(\ref{S-RGE}), $n_{\tilde{F}}$ represents degrees of freedom for $\tilde{F}$.
The solutions of (\ref{mF-RGE}) are written by
\begin{eqnarray}
m_{\tilde{F}}^{2}(Q) &=& m_{\tilde{F}}^{2}(M_U) 
 + \sum_{i=1}^{3}\frac{2 C_2^{(i)}(\tilde{F})}{b_{i}}\left(M_{i}^{2}(M_U)-M_{i}^{2}(Q)\right) 
\nonumber\\
&~& ~~~ +\frac{3}{5b_{1}}Y(\tilde{F})\left(S(Q)-S(M_U)\right)  
\nonumber \\
&=& m_{\tilde{F}}^{2}(M_U)+\sum_{i}\xi_{i}(\tilde{F})M_{i}^{2}(Q) +\xi_{S}(\tilde{F})S(Q) ,
\label{mF-sol}
\end{eqnarray}
where the coefficients $\xi_i(k)$ and $\xi_S(k)$ are defined by
\begin{eqnarray}
\xi_{i}(\tilde{F}) \equiv \frac{2 C_2^{(i)}(\tilde{F})}{b_{i}}\left(\left(\frac{\alpha_{i}(M_U)}{\alpha_{i}(Q)}\right)^{2}-1\right) ,~
\xi_{S}(\tilde{F}) \equiv \frac{3Y(\tilde{F})}{5b_{1}}\left(1-\frac{\alpha_{1}(M_U)}{\alpha_{1}(Q)}\right) .
\label{xiS}
\end{eqnarray}
Here we use the solution of $S(Q)$ given by
\begin{eqnarray}
S(Q) = \frac{\alpha_{1}(Q)}{\alpha_{1}(M_U)}S(M_U) .
\label{S-sol}
\end{eqnarray}
Values of $m_{\tilde{F}}^{2}(M_U)$ can be determined by using the solutions (\ref{mF-sol}) and the value $\alpha_i(M_U)$. 
If we do not know the value of $M_U$, we can use one of solutions to fix it.

For the third generation sfermions and Higgs doublets,\footnote{
Hereafter we add Higgs bosons ($H_1$, $H_2$) to $\tilde{F}$.} the relevant RGEs of their mass are given by 
\begin{eqnarray}
&~& \frac{{\rm d}\ }{{\rm d}t}m_{\tilde{F}}^{2} = 4 \sum_{i=1}^{3}C_2^{(i)}(\tilde{F}) \alpha_{i}M_{i}^{2}
 -\frac{3}{5}Y(\tilde{F})\alpha_{1}S+ {\mathcal Y}_{\tilde{F}} ,
\label{mF3-RGE}
\end{eqnarray}
where the ${\mathcal Y}_{\tilde{F}}$ stands for contributions from Yukawa interactions such that
\begin{eqnarray}
\hspace{-14mm}&~& {\mathcal Y}_{\tilde{q}_3}  = - \frac{f_t^2}{4\pi}\left(m_{\tilde{q}_3}^2 + m_{\tilde{t}^*_R}^2 + m_{H_2}^2 + A_t^2\right) 
 - \frac{f_b^2}{4\pi}\left(m_{\tilde{q}_3}^2 + m_{\tilde{b}^*_R}^2 + m_{H_1}^2 + A_b^2\right), 
\label{Yq3} \\
\hspace{-14mm}&~& {\mathcal Y}_{\tilde{t}^*_R}  = - \frac{f_t^2}{2\pi}\left(m_{\tilde{q}_3}^2 + m_{\tilde{t}^*_R}^2 + m_{H_2}^2 + A_t^2\right),
\label{YtR} \\
\hspace{-14mm}&~& {\mathcal Y}_{\tilde{b}^*_R}  = - \frac{f_b^2}{2\pi}\left(m_{\tilde{q}_3}^2 + m_{\tilde{b}^*_R}^2 + m_{H_1}^2 + A_b^2\right),
\label{YtR} \\
\hspace{-14mm}&~& {\mathcal Y}_{\tilde{l}_3}  =  - \frac{f_{\tau}^2}{4\pi}\left(m_{\tilde{l}_3}^2 
 + m_{\tilde{\tau}^*_R}^2 + m_{H_1}^2 + A_{\tau}^2\right) , 
\label{Yl3} \\
\hspace{-14mm}&~& {\mathcal Y}_{\tilde{\tau}^*_R}  = - \frac{f_{\tau}^2}{2\pi}\left(m_{\tilde{l}_3}^2 
 + m_{\tilde{\tau}^*_R}^2 + m_{H_1}^2 + A_{\tau}^2\right),
\label{YtauR} \\
\hspace{-14mm}&~& {\mathcal Y}_{H_1}  = - \frac{f_{\tau}^2}{4\pi}\left(m_{\tilde{l}_3}^2 + m_{\tilde{\tau}^*_R}^2 + m_{H_1}^2 + A_{\tau}^2\right) 
 - \frac{3f_b^2}{4\pi}\left(m_{\tilde{q}_3}^2 + m_{\tilde{b}^*_R}^2 + m_{H_1}^2 + A_b^2\right), 
\label{YH1} \\
\hspace{-14mm}&~& {\mathcal Y}_{H_2}  = - \frac{3f_t^2}{4\pi}\left(m_{\tilde{q}_3}^2 + m_{\tilde{t}^*_R}^2 + m_{H_2}^2 + A_t^2\right), 
\label{YH2} 
\end{eqnarray}
where $f_t$, $f_b$ and $f_{\tau}$ are Yukawa couplings. 
The solutions of (\ref{mF3-RGE}) are formally written by\footnote{
The exact analytical expressions for ${\mathcal F}_{\tilde{F}}$ have not been known in case with whole third generation Yukawa couplings.}
\begin{eqnarray}
m_{\tilde{F}}^{2}(Q) = m_{\tilde{F}}^{2}(M_U)+\sum_{i}\xi_{i}(\tilde{F})M_{i}^{2}(Q) +\xi_{S}(\tilde{F})S(Q)+ {\mathcal F}_{\tilde{F}} ,
\label{mF3-sol}
\end{eqnarray}
where $d{\mathcal F}_{\tilde{F}}/dt={\mathcal Y}_{\tilde{F}}$.
By using RG Eq. (\ref{mF3-RGE}) and RG Eqs. of Yukawa couplings and $A$ parameters, 
values of ${\mathcal F}_{\tilde{F}}$ and $m_{\tilde{F}}^{2}(M_U)$ can be determined numerically.

\subsection{Sparticle sum rules -- general arguments}

In the previous subsection, we encounter several sum rules among sparticles and Higgs bosons, which are 
irrelevant to the structure of models beyond the MSSM, and can be the touchstone of MSSM.
In this subsection, we give general arguments on sum rules among sparticles which are relevant to physics beyond the MSSM.

First we consider gaugino masses. 
We assume that gaugino masses take the following values at some high energy scale $M_U$: 
\begin{eqnarray}
M_{i}(M_U) = l_{i} M_{1/2} ,
\label{Mi0}
\end{eqnarray}
where $l_i$s are some constant numbers and $M_{1/2}$ is a mass parameter.
Then the values of $M_i$ at $M_{EW}$ are given by
\begin{eqnarray}
M_{i}(M_{EW}) = l_i \frac{\alpha_i(M_{EW})}{\alpha_i(M_U)}M_{1/2} 
 = l_{i}\left(1-\frac{b_{i}}{2\pi}\alpha_{i}\ln \frac{M_U}{M_{EW}}\right)M_{1/2} .
\label{Mi-EW} 
\end{eqnarray}
Hereafter we abbreviate $M_i(M_{EW})$ and $\alpha_i(M_{EW})$ as $M_i$ and $\alpha_i$, respectively.
Same applies to $M_{\tilde{f}}^2$ and $m_f^2$.
Using Eq.(\ref{Mi-EW}), we obtain the sum rule,
\begin{eqnarray}
\frac{\alpha_{i}(M_U)}{l_i \alpha_{i}}M_i = \frac{\alpha_{j}(M_U)}{l_j \alpha_{j}}M_j  .
\label{Mi-SR}
\end{eqnarray}
As a simple but interesting case, we consider the case with a universal gaugino mass, i.e., $M_{i}(M_U) = M_{1/2}$,
and the gauge coupling unification, i.e., $\alpha_i(M_U) = \alpha_U$.
In this case, we obtain so-called GUT-relation,
\begin{eqnarray}
\frac{M_1}{\alpha_{1}} = \frac{M_2}{\alpha_{2}} = \frac{M_3}{\alpha_{3}} = \frac{M_{1/2}}{\alpha_U} .
\label{GUT-relation}
\end{eqnarray}
This relation holds on at the 1-loop level even when the gauge symmetry breaks to the SM one in several steps, if
the gauge group is unified into a simple group at some high energy scale.\cite{KM&Y}

Next we consider scalar masses.
We assume that there exist the following $n$ kinds of relations among scalar mass parameters at $M_U$:
\begin{eqnarray}
\sum_{\tilde{F}} a_{\tilde{F}}^{(k)} m_{\tilde{F}}^2(M_U) = 0  ,
\label{SRmu0}
\end{eqnarray} 
where $a_{\tilde{F}}^{(k)}$ ($k = 1, \cdots, n$) are some constants.
Then the following specific sum rules are derived:
\begin{eqnarray}
&~& a_{H_1}^{(k)} m_{H_1}^2 + a_{H_2}^{(k)} m_{H_2}^2 
 + \sum_{\tilde{f}} a_{\tilde{F}}^{(k)} M_{\tilde{f}}^2 - \sum_{f} a_{\tilde{F}}^{(k)} m_f^2 
  - \sum_{\tilde{f}} a_{\tilde{F}}^{(k)} D_{W}(\tilde{f})
\nonumber \\
&~& ~~~ - \sum_{\tilde{F}} a_{\tilde{F}}^{(k)} \sum_i \xi_i(\tilde{F}) M_i^2 - \sum_{\tilde{F}} a_{\tilde{F}}^{(k)} \xi_S(\tilde{F}) S 
 - \sum_{\tilde{F}} a_{\tilde{F}}^{(k)} {\mathcal F}_{\tilde{F}} = 0  
\label{Mf-SR}
\end{eqnarray}
by using Eqs.(\ref{Mf}), (\ref{mF-sol}) and (\ref{mF3-sol}).
If the GUT-relation (\ref{GUT-relation}) holds on, the above sum rules are rewritten by
\begin{eqnarray}
\hspace{-10mm} &~& a_{H_1}^{(k)} m_{H_1}^2 + a_{H_2}^{(k)} m_{H_2}^2 
 + \sum_{\tilde{f}} a_{\tilde{F}}^{(k)} M_{\tilde{f}}^2 - \sum_{f} a_{\tilde{F}}^{(k)} m_f^2 
  - \sum_{\tilde{f}} a_{\tilde{F}}^{(k)} D_{W}(\tilde{f})
\nonumber \\
\hspace{-10mm} &~& ~~~ - \sum_{\tilde{F}} a_{\tilde{F}}^{(k)} \sum_i \left(\frac{\alpha_i}{\alpha_3}\right)^2 \xi_i(\tilde{F}) M_{\tilde{g}}^2
 - \sum_{\tilde{F}} a_{\tilde{F}}^{(k)} \xi_S(\tilde{F}) S 
 - \sum_{\tilde{F}} a_{\tilde{F}}^{(k)} {\mathcal F}_{\tilde{F}} = 0 . 
\label{Mf-SR-GUT}
\end{eqnarray}
The (\ref{Mf-SR}) or (\ref{Mf-SR-GUT}) is the master formula of our analysis.


We discuss the case with anomaly mediation\cite{anomaly}, in which the hypothesis of desert violates on the face of it. 
When we take effects of anomaly mediation in SUGRA into account,
gaugino masses are given by 
\begin{eqnarray}
M_{i}(Q) = \left(1- \frac{b_{i}}{2\pi}\alpha_{i}\ln\frac{M_U}{Q}\right)M_{1/2} +\frac{b_{i}}{4\pi}\alpha_{i}m_{3/2} ,
\label{Mi-anomaly}
\end{eqnarray}
where the second term represents the contribution from anomaly mediation.
The first one is an ordinary contribution and here we take a common gaugino mass $M_{1/2}$.
The above formula is rewritten as
\begin{eqnarray}
M_{i} = \left(1-\frac{b_{i}}{2\pi}\alpha_{i}\ln\frac{Q_{\ast}}{M_{EW}}\right) M_{1/2} = \frac{\alpha_{i}}{\alpha_{i}(Q_{\ast})}M_{1/2},
\end{eqnarray}
where $Q_{\ast}$ is a new scale defined by 
\begin{eqnarray}
Q_{\ast} \equiv \frac{M_U}{\left({M_U}/{m_{3/2}}\right)^{x}} ,~~
 x \equiv \frac{m_{3/2}}{2M_{1/2}\ln({M_U}/{m_{3/2}})}.
\end{eqnarray}
At the scale $Q_*$, gaugino masses unify, but it does not necessarily mean the existence of new physics there.
Hence the unification is regarded as a mirage and this type of SUSY breaking mediation is called $\lq$mirage mediation'.\cite{mirage}
If some conditions for parameters fulfills, this kind of unification occurs for scalar masses.
It is possible to select models beyond the MSSM by checking sum rules such as (\ref{GUT-relation}) and (\ref{Mf-SR-GUT}).
Using one of sum rules, the scale $M_U$ is determined and other sum rules are tested.
If the value of $M_U$ is not an expected one from a theoretical prediction, there is a possibility that mirage mediation occurs.

\subsection{Sparticle sum rules -- examples}

We derive specific sum rules among scalar particles taking two simple examples such as a universal type of 
soft SUSY breaking scalar masses at $M_U$ and a generation dependent type there. 
Former one has been intensively studied in Ref.\cite{FHK&N,M&R}.
Here we study more generic situation with the case that $\tan\beta$ is large, i.e., the bottom and tau Yukawa couplings are not
negligible.

\subsubsection{Universal type}

Let us discuss the case with a universal type of soft SUSY breaking scalar masses at $M_U$, 
i.e., $m_{\tilde{F}}^2(M_U) = m_0^2$.\footnote{
In the SUGRA, $M_U$ is regarded as the gravitational scale $M \equiv M_{pl}/\sqrt{8\pi}$.
Such a mass degeneracy originates from the minimal SUGRA\cite{mSUGRA} or, in general, the SUGRA 
which possesses $SU(n)$ symmetry in the K\"ahler potential concerning $n$ kinds of matter multiplets.\cite{HL&W}}
By using the mass formula (\ref{Mf}), the solutions (\ref{mF-sol}) and (\ref{mF3-sol}), the scalar masses at $M_{EW}$ are given by
\begin{eqnarray}
&~& M_{\tilde{u}_{L}}^{2} = M_{\tilde{c}_{L}}^{2} 
= m_{0}^{2} + \zeta_3 M_3^2 + \zeta_2 M_2^2 + \zeta_1 M_1^2 
\nonumber \\
&~& ~~~~~~~~ + \left(\frac{2}{3}M_{W}^{2} - \frac{1}{6}M_{Z}^{2}\right)\cos 2\beta , \\
&~& M_{\tilde{d}_{L}}^{2} = M_{\tilde{s}_{L}}^{2} 
= m_{0}^{2} + \zeta_3 M_3^2 + \zeta_2 M_2^2 + \zeta_1 M_1^2 
\nonumber \\
&~& ~~~~~~~~ + \left(-\frac{1}{3}M_{W}^{2} - \frac{1}{6}M_{Z}^{2}\right)\cos 2\beta , \\
&~& M_{\tilde{u}_{R}}^{2} = M_{\tilde{c}_{R}}^{2} 
= m_{0}^{2} +\zeta_3 M_3^2 +16\zeta_1 M_1^2 + \left(-\frac{2}{3}M_{W}^{2}+\frac{2}{3}M_{Z}^{2}\right)\cos 2\beta , ~~~\\
&~& M_{\tilde{d}_{R}}^{2} = M_{\tilde{s}_{R}}^{2} 
= m_{0}^{2} + \zeta_3 M_3^2 + 4\zeta_1 M_1^2 +\left(\frac{1}{3}M_{W}^{2} -\frac{1}{3}M_{Z}^{2}\right)\cos 2\beta , \\
&~& M_{\tilde{\nu}_{eL}}^{2} = M_{\tilde{\nu}_{\mu L}}^{2} 
= m_{0}^{2} + \zeta_2 M_2^2 + 9\zeta_1 M_1^2 +\frac{1}{2}M_{Z}^{2}\cos 2\beta , \\
&~& M_{\tilde{e}_{L}}^{2} = M_{\tilde{\mu}_{L}}^{2} 
= m_{0}^{2}+\zeta_2 M_2^2+9\zeta_1 M_1^2 +\left(-M_{W}^{2}+\frac{1}{2}M_{Z}^{2}\right)\cos 2\beta , \\
&~& M_{\tilde{e}_{R}}^{2} = M_{\tilde{\mu}_{R}}^{2} 
= m_{0}^{2} +36\zeta_1 M_1^2 +\left(M_{W}^{2}-M_{Z}^{2}\right)\cos 2\beta , \\
&~& M_{\tilde{t}_{L}}^{2} 
= m_{0}^{2} + \zeta_3 M_3^2+\zeta_2 M_2^2+\zeta_3 M_3^2 + \left(\frac{2}{3}M_{W}^{2} - \frac{1}{6}M_{Z}^{2}\right)\cos 2\beta 
\nonumber \\
&~& ~~~~~~~~ -F_{t} -F_{b} +m_{t}^{2} , \\
&~& M_{\tilde{b}_{L}}^{2} 
= m_{0}^{2} + \zeta_3 M_3^2 + \zeta_2 M_2^2 +\zeta_1 M_1^2 + \left(-\frac{1}{3}M_{W}^{2} - \frac{1}{6}M_{Z}^{2}\right)\cos 2\beta 
\nonumber \\
&~& ~~~~~~~~ -F_{t} -F_{b} +m_{b}^{2} , \\
&~& M_{\tilde{t}_{R}}^{2} 
= m_{0}^{2} +\zeta_3 M_3^2+16\zeta_1 M_1^2 + \left(-\frac{2}{3}M_{W}^{2}+\frac{2}{3}M_{Z}^{2}\right)\cos 2\beta 
\nonumber \\
&~& ~~~~~~~~ -2F_{t}+m_{t}^{2} , \\
&~& M_{\tilde{b}_{R}}^{2} 
= m_{0}^{2} + \zeta_3 M_3^2 + 4\zeta_1 M_1^2 +\left(\frac{1}{3}M_{W}^{2} -\frac{1}{3}M_{Z}^{2}\right)\cos 2\beta
\nonumber \\
&~& ~~~~~~~~ -2F_{b} +m_{b}^{2} , \\
&~& M_{\tilde{\nu}_{\tau L}}^{2} 
= m_{0}^{2} + \zeta_2 M_2^2+9\zeta_1 M_1^2+\frac{1}{2}M_{Z}^{2}\cos 2\beta-F_{\tau} , \\
&~& M_{\tilde{\tau}_{L}}^{2} 
= m_{0}^{2}+\zeta_2 M_2^2+9\zeta_1 M_1^2 +\left(-M_{W}^{2}+\frac{1}{2}M_{Z}^{2}\right)\cos 2\beta
\nonumber \\
&~& ~~~~~~~~ -F_{\tau} +m_{\tau}^{2} , \\
&~& M_{\tilde{\tau}_{R}}^{2} 
= m_{0}^{2} +36\zeta_1 M_1^2+\left(M_{W}^{2}-M_{Z}^{2}\right)\cos 2\beta -2F_{\tau} +m_{\tau}^{2} , \\
&~& m_{H_1}^2 
= m_{0}^{2} + \zeta_2 M_2^2+9\zeta_1 M_1^2-F_{\tau}-3F_b , \\
&~& m_{H_2}^2 
= m_{0}^{2} + \zeta_2 M_2^2+9\zeta_1 M_1^2-3F_t ,
\end{eqnarray}
where we neglect effects of Yukawa couplings in the first two generations and
we use $S = 0$.
Hereafter we neglect $m_b$ and $m_{\tau}$ for simplicity.
The $\zeta_3$, $\zeta_2$ and $\zeta_1$ are defined by
\begin{eqnarray}
&~& \zeta_3 = -\frac{8}{9}\left(\left(\frac{\alpha_{3}(M_U)}{\alpha_{3}}\right)^{2}-1\right) , ~~
 \zeta_2 = \frac{3}{2}\left(\left(\frac{\alpha_{2}(M_U)}{\alpha_{2}}\right)^{2}-1\right) , 
\nonumber \\
&~& \zeta_1 = \frac{1}{198}\left(\left(\frac{\alpha_{1}(M_U)}{\alpha_{1}}\right)^{2}-1\right) .
\end{eqnarray}
The $F_{t}$, $F_b$ and $F_{\tau}$ stands for effects of Yukawa interactions in the third generation
and they satisfy the following equations,
\begin{eqnarray}
&~& \frac{dF_t}{dt} = \frac{f_t^2}{4\pi}\left(m_{\tilde{q}_3}^2 + m_{\tilde{t}_R}^2 + m_{H_2}^2 + A_t^2\right) , 
\label{Ft} \\
&~& \frac{dF_b}{dt} = \frac{f_b^2}{4\pi}\left(m_{\tilde{q}_3}^2 + m_{\tilde{b}_R}^2 + m_{H_1}^2 + A_b^2\right) , 
\label{Fb} \\
&~& \frac{dF_{\tau}}{dt} = \frac{f_{\tau}^2}{4\pi}\left(m_{\tilde{l}_3}^2 + m_{\tilde{\tau}_R}^2 + m_{H_1}^2 + A_{\tau}^2\right) . 
\label{Ftau} 
\end{eqnarray}
Values of $F_t$, $F_b$ and $F_{\tau}$ are determined numerically by solving RG Eqs. of sparticle masses and coupling constants.

Twenty-two kinds of sum rules must exist because 
there are twenty-three kinds of scalar mass parameters 
($M_{\tilde{u}_L}^2$, $M_{\tilde{u}_L}^2$, $\cdots$, $M_{\tilde{\tau}_R}^2$, $m_{H_1}^2$, $m_{H_2}^2$)
and one unknown parameter $m_0$.
By eliminating $m_{0}$, we, in fact, obtain the following twenty-two sum rules, 
\begin{eqnarray}
&~& M_{\tilde{u}_{L}}^{2}-M_{\tilde{d}_{L}}^{2} = M_{\tilde{c}_{L}}^{2}-M_{\tilde{s}_{L}}^{2} 
= M_{\tilde{t}_{L}}^{2}-M_{\tilde{b}_{L}}^{2} - m_t^2 = M_{W}^{2}\cos 2\beta ,
\label{SR1} \\
&~& M_{\tilde{\nu}_{eL}}^{2}-M_{\tilde{e}_{L}}^{2} = M_{\tilde{\nu}_{\mu L}}^{2}-M_{\tilde{\mu}_{L}}^{2} 
= M_{\tilde{\nu}_{\tau L}}^{2}-M_{\tilde{\tau}_{L}}^{2} = M_{W}^{2}\cos 2\beta ,
\label{SR2} \\
&~& M_{\tilde{u}_{L}}^{2}-M_{\tilde{u}_{R}}^{2} = M_{\tilde{c}_{L}}^{2}-M_{\tilde{c}_{R}}^{2} 
= M_{\tilde{t}_{L}}^{2}-M_{\tilde{t}_{R}}^{2} - F_t + F_b 
\nonumber \\
&~& ~~~ = \zeta_2 M_2^2-15\zeta_1 M_1^2+\left(\frac{4}{3}M_{W}^{2}-\frac{5}{6}M_{Z}^{2}\right)\cos 2\beta , 
\label{SR3} \\
&~& M_{\tilde{d}_{L}}^{2}-M_{\tilde{e}_{R}}^{2} = M_{\tilde{s}_{L}}^{2}-M_{\tilde{\mu}_{R}}^{2} 
= M_{\tilde{b}_{L}}^{2}-M_{\tilde{\tau}_{R}}^{2} + F_t + F_b - 2 F_{\tau}
\nonumber \\
&~& ~~~ = \zeta_3 M_3^2+\zeta_2 M_2^2-35\zeta_1 M_1^2+\left(-\frac{4}{3}M_{W}^{2}+\frac{5}{6}M_Z^2\right)\cos 2\beta ,
\label{SR4} \\
&~& M_{\tilde{e}_{L}}^{2}-M_{\tilde{d}_{R}}^{2} = M_{\tilde{\mu}_{L}}^{2}-M_{\tilde{s}_{R}}^{2} 
= M_{\tilde{\tau}_{L}}^{2}-M_{\tilde{b}_{R}}^{2} + F_{\tau} - 2 F_b
\nonumber \\
&~& ~~~ = -\zeta_3 M_3^2+\zeta_2 M_2^2+5\zeta_1 M_1^2+\left(-\frac{4}{3}M_{W}^{2}+\frac{5}{6}M_Z^2\right)\cos 2\beta ,
\label{SR5} \\
&~& M_{\tilde{u}_{R}}^{2}-M_{\tilde{d}_{R}}^{2} = M_{\tilde{c}_{R}}^{2}-M_{\tilde{s}_{R}}^{2} 
= M_{\tilde{t}_{R}}^{2}-M_{\tilde{b}_{R}}^{2} - m_t^2 + 2F_t - 2F_b 
\nonumber \\
&~& ~~~ = 12\zeta_1 M_1^2+\left(-M_{W}^{2}+M_{Z}^{2}\right)\cos 2\beta ,
\label{SR6}\\
&~& M_{\tilde{u}_{L}}^{2} = M_{\tilde{c}_{L}}^{2} = M_{\tilde{t}_{L}}^{2} -m_{t}^{2} +F_{t}+F_{b} ,
\label{SR7}\\
&~& m_{H_{1}}^{2}-m_{H_{2}}^{2} = -F_{\tau} -3F_{b} + 3F_{t} ,
\label{H1}\\
&~& m_{H_{2}}^{2} = M_{e_{L}}^{2} + \left(M_{W}^{2}-\frac{1}{2}M_{Z}^{2}\right)\cos 2\beta -3F_{t} .
\label{H2}
\end{eqnarray}
They are also expressed as
\begin{eqnarray}
&~& M_{\tilde{u}_{L}}^{2}-M_{\tilde{d}_{L}}^{2} = M_{W}^{2}\cos 2\beta ,
\label{SR1-1} \\
&~& M_{\tilde{\nu}_{eL}}^{2}-M_{\tilde{e}_{L}}^{2} = M_{W}^{2}\cos 2\beta ,
\label{SR1-2} \\
&~& M_{\tilde{u}_{L}}^{2}-M_{\tilde{u}_{R}}^{2} = \zeta_2 M_2^2-15\zeta_1 M_1^2
+\left(\frac{4}{3}M_{W}^{2}-\frac{5}{6}M_{Z}^{2}\right)\cos 2\beta , 
\label{SR1-3} \\
&~& M_{\tilde{d}_{L}}^{2}-M_{\tilde{e}_{R}}^{2} = \zeta_3 M_3^2+\zeta_2 M_2^2-35\zeta_1 M_1^2
\nonumber \\
&~& ~~~~~~~~~~~~~~~~~ +\left(-\frac{4}{3}M_{W}^{2}+\frac{5}{6}M_Z^2\right)\cos 2\beta ,
\label{SR1-4} \\
&~& M_{\tilde{e}_{L}}^{2}-M_{\tilde{d}_{R}}^{2} = -\zeta_3 M_3^2+\zeta_2 M_2^2+5\zeta_1 M_1^2
\nonumber \\
&~& ~~~~~~~~~~~~~~~~ +\left(-\frac{4}{3}M_{W}^{2}+\frac{5}{6}M_Z^2\right)\cos 2\beta ,
\label{SR1-5} \\
&~& M_{\tilde{u}_{R}}^{2}-M_{\tilde{d}_{R}}^{2} = 12\zeta_1 M_1^2+\left(-M_{W}^{2}+M_{Z}^{2}\right)\cos 2\beta ,
\label{SR1-6}\\
&~& M_{\tilde{u}_{L}}^{2} = M_{\tilde{c}_{L}}^{2} = M_{\tilde{t}_{L}}^{2} -m_{t}^{2} +F_{t}+F_{b} ,
\label{SRu-L} \\
&~& M_{\tilde{d}_{L}}^{2} = M_{\tilde{s}_{L}}^{2} = M_{\tilde{b}_{L}}^{2} +F_{t}+F_{b} ,
\label{SRd-L} \\
&~& M_{\tilde{u}_{R}}^{2} = M_{\tilde{c}_{R}}^{2} = M_{\tilde{t}_{R}}^{2} -m_{t}^{2} +2F_{t} ,
\label{SRu-R} \\
&~& M_{\tilde{d}_{R}}^{2} = M_{\tilde{s}_{R}}^{2} = M_{\tilde{b}_{L}}^{2}  +2F_{b} ,
\label{SRd-R} \\
&~& M_{\tilde{\nu}_{eL}}^{2} = M_{\tilde{\nu}_{\mu L}}^{2} = M_{\tilde{\nu}_{\tau L}}^{2} +F_{\tau} ,
\label{SRnu-L} \\
&~& M_{\tilde{e}_{L}}^{2} = M_{\tilde{\mu}_{L}}^{2} = M_{\tilde{\tau}_{L}}^{2}+F_{\tau} ,
\label{SRe-L}\\
&~& M_{\tilde{e}_{R}}^{2} = M_{\tilde{\mu}_{R}}^{2} = M_{\tilde{\tau}_{R}}^{2}  +2F_{\tau} ,
\label{SRe-R}\\
&~& m_{H_{1}}^{2}-m_{H_{2}}^{2} = -F_{\tau} -3F_{b} + 3F_{t} ,
\label{H1}\\
&~& m_{H_{2}}^{2} = M_{e_{L}}^{2} + \left(M_{W}^{2}-\frac{1}{2}M_{Z}^{2}\right)\cos 2\beta -3F_{t} .
\label{H2}
\end{eqnarray}
If all paremeters were measured precisely enough, these sum rules can be powerful tools to test the universality of
scalar mass parameters at $M_U$.
If any parameters have ambiguities, we should use sum rules in which such parameters are absent.
Here we discuss three typical cases that parameters are eliminated.

~~\\
(i) Elimination of  $\cos 2\beta$

In this case, there exist twenty-one sum rules, i.e., fifteen relations (\ref{SRu-L}) - (\ref{H1}) and
six relations such that
\begin{eqnarray}
\hspace{-10mm} &~& M_{\tilde{u}_{L}}^{2}-M_{\tilde{d}_{L}}^{2} = M_{\tilde{\nu}_{eL}}^{2} - M_{\tilde{e}_{L}}^{2} , \\
\hspace{-10mm} &~& M_{\tilde{u}_{L}}^{2}-M_{\tilde{u}_{R}}^{2} +M_{\tilde{d}_{L}}^{2} - M_{\tilde{e}_{R}}^{2} 
 = \zeta_3 M_3^2+2\zeta_2 M_2^2-50\zeta_1 M_1^2 , \\
\hspace{-10mm} &~& M_{\tilde{d}_{L}}^{2}-M_{\tilde{e}_{R}}^{2} -M_{\tilde{e}_{L}}^{2} + M_{\tilde{d}_{R}}^{2} = 2\zeta_3 M_3^2-40\zeta_1 M_1^2 , \\
\hspace{-10mm} &~& 3 \left(M_{\tilde{\nu}_{e}}^{2}+M_{\tilde{e}_{L}}^{2}\right) -M_{\tilde{d}_{R}}^{2} - 5M_{\tilde{u}_{R}}^{2} 
 = -6\zeta_3 M_3^2+6\zeta_2 M_2^2-30\zeta_1 M_1^2 , \\
\hspace{-10mm} &~& M_{\tilde{u}_{R}}^{2}-M_{\tilde{d}_{R}}^{2} 
= 12\zeta_1 M_1^2 -\left(1-\frac{M_{Z}^{2}}{M_{W}^{2}}\right)\left(M_{\tilde{u}_{L}}^{2}-M_{\tilde{d}_{L}}^{2}\right) , \\
\hspace{-10mm} &~& m_{H_{2}}^{2} = M_{e_{L}}^{2} + \left(1-\frac{M_{Z}^{2}}{2M_{W}^{2}}\right)\left(M_{\tilde{u}_{L}}^{2}-M_{\tilde{d}_{L}}^{2}\right) 
 -3F_{t} .
\end{eqnarray}

~~\\
(ii) Elimination of $\zeta_i$s

In this case, there exist nineteen sum rules, i.e., eighteen relations (\ref{SR1-1}) , (\ref{SR1-2}), (\ref{SRu-L}) - (\ref{H2}) and
one relation such that
\begin{eqnarray}
2M_{\tilde{u}_{R}}^{2}-M_{\tilde{d}_{R}}^{2}-M_{\tilde{d}_{L}}^{2}+M_{\tilde{e}_{L}}^{2}-M_{\tilde{e}_{R}}^{2} 
 = \frac{10}{3}\left(M_{Z}^{2}-M_{W}^{2}\right)\cos 2\beta .
\end{eqnarray}

~~\\
(iii) Elimination of $F_t$, $F_b$, $F_{\tau}$, $A_t$, $A_b$ and $A_{\tau}$

In this case, there exist sixteen sum rules, i.e., thirteen relations such as
the first and second generation part of (\ref{SR1}) - (\ref{SR7}), and
three sum rules such that
\begin{eqnarray}
&~& M_{\tilde{\tau}_1}^2 + M_{\tilde{\tau}_2}^2 - 3 M_{\tilde{\nu}_{\tau}}^2 
= M_{\tilde{e}_L}^2 + M_{\tilde{e}_R}^2 - 3 M_{\tilde{\nu}_{eL}}^2 ,
\label{SR-a} \\
&~& 3\left(M_{\tilde{t}_1}^2 + M_{\tilde{t}_2}^2 - M_{\tilde{b}_1}^2 - M_{\tilde{b}_2}^2\right) 
 + 2\left(m_{H_1}^2 - m_{H_2}^2 - M_{\tilde{\nu}_{\tau L}}^2\right) 
\nonumber \\
&~& ~~~ = 3\left(M_{\tilde{u}_L}^2 + M_{\tilde{u}_R}^2 + 2 m_t^2 - M_{\tilde{d}_L}^2 - M_{\tilde{d}_R}^2\right) - 2M_{\tilde{\nu}_{eL}}^2 ,
\label{SR-b} \\ 
&~& 3\left(M_{\tilde{t}_1}^2 + M_{\tilde{t}_2}^2 + M_{\tilde{b}_1}^2 + M_{\tilde{b}_2}^2\right) 
 -4\left(m_{H_1}^2 + m_{H_2}^2 - M_{\tilde{\nu}_{\tau L}}^2\right)  
\nonumber \\
&~& ~~~ = 3\left(M_{\tilde{u}_L}^2 + M_{\tilde{u}_R}^2 + 2 m_t^2 + M_{\tilde{d}_L}^2 + M_{\tilde{d}_R}^2\right) 
\nonumber \\
&~& ~~~~~~ - 8 M_{\tilde{e}_{L}}^2 + 4M_{\tilde{\nu}_{eL}}^2 -\left(8M_W^2 - 4M_Z^2\right)\cos 2\beta .
\label{SR-c}
\end{eqnarray}

\subsubsection{Generation dependent type}

We study the case that soft SUSY breaking scalar masses are generation dependent ones at $M_U$, i.e.,
there are five independent mass parameters such that
\begin{eqnarray}
\hspace{-10mm} &~& {m}_{\tilde{q}_1}^{2}(M_U) = {m}_{\tilde{u}^*_R}^{2}(M_U) = {m}_{\tilde{d}^*_R}^{2}(M_U)
= {m}_{\tilde{l}_{1}}^{2}(M_U) = {m}_{\tilde{e}^*_R}^{2}(M_U) \equiv {m}_{0}^{(1)2} ,~ 
\label{m-1} \\
\hspace{-10mm} &~& {m}_{\tilde{q}_2}^{2}(M_U) = {m}_{\tilde{c}^*_R}^{2}(M_U) = {m}_{\tilde{s}^*_R}^{2}(M_U)
= {m}_{\tilde{l}_{2}}^{2}(M_U) = {m}_{\tilde{\mu}^*_R}^{2}(M_U) \equiv {m}_{0}^{(2)2} ,~ 
\label{m-2} \\
\hspace{-10mm} &~& {m}_{\tilde{q}_3}^{2}(M_U) = {m}_{\tilde{t}^*_R}^{2}(M_U) = {m}_{\tilde{b}^*_R}^{2}(M_U)
= {m}_{\tilde{l}_{3}}^{2}(M_U) = {m}_{\tilde{\tau}^*_R}^{2}(M_U) \equiv {m}_{0}^{(3)2} ,~ 
\label{m-3} \\
\hspace{-10mm} &~& m_{H_1}^2(M_U) \equiv m_{0}^{(H_{1})2} ,~~ m_{H_2}^2(M_U) \equiv m_{0}^{(H_{2})2} .  
\end{eqnarray}
Note that $S(M_U) = m_{H_2}^2(M_U) - m_{H_1}^2(M_U)$ and hence $S$ is non-vanishing if $m_{H_1}^2(M_U) \neq m_{H_2}^2(M_U)$.
In this case, we have eighteen sum rules, i.e., six relations (\ref{SR1}) and (\ref{SR2}), and twelve ones
\begin{eqnarray}
\hspace{-7mm}&~& M_{\tilde{u}_{L}}^{2}-M_{\tilde{u}_{R}}^{2} = M_{\tilde{c}_{L}}^{2}-M_{\tilde{c}_{R}}^{2} 
= M_{\tilde{t}_{L}}^{2}-M_{\tilde{t}_{R}}^{2} - F_t + F_b 
\nonumber \\
\hspace{-7mm}&~& ~~~ = \zeta_2 M_2^2-15\zeta_1 M_1^2+\left(\frac{4}{3}M_{W}^{2}-\frac{5}{6}M_{Z}^{2}\right)\cos 2\beta +5 \mathcal{S} , 
\label{SR3K} \\
\hspace{-7mm}&~& M_{\tilde{d}_{L}}^{2}-M_{\tilde{e}_{R}}^{2} = M_{\tilde{s}_{L}}^{2}-M_{\tilde{\mu}_{R}}^{2} 
= M_{\tilde{b}_{L}}^{2}-M_{\tilde{\tau}_{R}}^{2} + F_t + F_b - 2 F_{\tau}
\nonumber \\
\hspace{-7mm}&~& ~~~ = \zeta_3 M_3^2+\zeta_2 M_2^2-35\zeta_1 M_1^2+\left(-\frac{4}{3}M_{W}^{2}+\frac{5}{6}M_Z^2\right)\cos 2\beta - 5 \mathcal{S} ,
\label{SR4K} \\
\hspace{-7mm}&~& M_{\tilde{e}_{L}}^{2}-M_{\tilde{d}_{R}}^{2} = M_{\tilde{\mu}_{L}}^{2}-M_{\tilde{s}_{R}}^{2} 
= M_{\tilde{\tau}_{L}}^{2}-M_{\tilde{b}_{R}}^{2} + F_{\tau} - 2 F_b
\nonumber \\
\hspace{-7mm}&~& ~~~ = -\zeta_3 M_3^2+\zeta_2 M_2^2+5\zeta_1 M_1^2+\left(-\frac{4}{3}M_{W}^{2}+\frac{5}{6}M_Z^2\right)\cos 2\beta - 5 \mathcal{S} ,
\label{SR5K} \\
\hspace{-5mm}&~& M_{\tilde{u}_{R}}^{2}-M_{\tilde{d}_{R}}^{2} = M_{\tilde{c}_{R}}^{2}-M_{\tilde{s}_{R}}^{2} 
= M_{\tilde{t}_{R}}^{2}-M_{\tilde{b}_{R}}^{2} -m_t^2 + 2F_t - 2F_b 
\nonumber \\
\hspace{-5mm}&~& ~~~ = 12\zeta_1 M_1^2+\left(-M_{W}^{2}+M_{Z}^{2}\right)\cos 2\beta - 6 \mathcal{S} ,
\label{SR6K}
\end{eqnarray}
where $\mathcal{S}$ is defined by
\begin{eqnarray}
\mathcal{S} \equiv \frac{1}{6}\frac{3}{5b_{1}}\left(1-\frac{\alpha_{1}(M_{U})}{\alpha_{1}}\right)S 
 = \frac{1}{10 b_{1}}\left(1-\frac{\alpha_{1}(M_{U})}{\alpha_{1}}\right) \left(m_{H_2}^2 - m_{H_1}^2\right) .
\label{mathcal-S}
\end{eqnarray}
In place of (\ref{SR3K}) or (\ref{SR4K}), we can use the following one,
\begin{eqnarray}
&~& M_{\tilde{u}_{R}}^{2} - M_{\tilde{e}_{R}}^{2} = M_{\tilde{c}_{R}}^{2} - M_{\tilde{\mu}_{R}}^{2}  
\nonumber \\
&~& ~~~ = M_{\tilde{t}_{R}}^{2} - M_{\tilde{\tau}_{R}}^{2} - m_t^2 + 2 F_t -2F_{\tau}
\nonumber \\
&~& ~~~ = \zeta_3 M_3^2-20\zeta_1 M_1^2+\left(-\frac{5}{3}M_{W}^{2}+\frac{5}{3}M_{Z}^{2}\right)\cos 2\beta -10 \mathcal{S} .
\label{SR4K'} 
\end{eqnarray}
By eliminating $\mathcal{S}$, we have following relations,
\begin{eqnarray}
\hspace{-15mm}&~& M_{\tilde{u}_{L}}^{2}-M_{\tilde{u}_{R}}^{2} + M_{\tilde{d}_{L}}^{2}-M_{\tilde{e}_{R}}^{2} 
 = \zeta_3 M_3^2 + 2 \zeta_2 M_2^2 - 50 \zeta_1 M_1^2 , 
\label{SK3noK} \\
\hspace{-15mm}&~& M_{\tilde{u}_{L}}^{2}-M_{\tilde{u}_{R}}^{2} + M_{\tilde{e}_{L}}^{2}-M_{\tilde{d}_{R}}^{2} 
 = - \zeta_3 M_3^2 + 2 \zeta_2 M_2^2 - 10 \zeta_1 M_1^2 , 
\label{SK4noK} \\
\hspace{-15mm}&~& 6\left(M_{\tilde{u}_{L}}^{2}-M_{\tilde{u}_{R}}^{2}\right) + 5\left(M_{\tilde{u}_{R}}^{2}-M_{\tilde{d}_{R}}^{2}\right) 
 = 6 \zeta_2 M_2^2 - 30 \zeta_1 M_1^2 + 3 M_W^2 \cos 2\beta .
\label{SK5noK}
\end{eqnarray}

If we require the mass degeneracy between the first two generations, i.e., ${m}_{0}^{(1)2} = {m}_{0}^{(2)2}$, 
we obtain extra sum rules $M_{\tilde{u}_{L}}^{2} = M_{\tilde{c}_{L}}^{2}$, 
$M_{\tilde{d}_{L}}^{2} = M_{\tilde{s}_{L}}^{2}$ and so on, and the sufficient suppression of FCNC processes is realized.

\section{Scalar sum rules in grand unification}

We derive peculiar sum rules among scalar masses in cases with
several kinds of non-universal soft SUSY breaking scalar mass parameters at $M_U$
based on SUSY GUTs in four and five dimensions.

\subsection{Grand unification in four dimension}

Scalar mass relations have been derived for various symmetry breaking patterns
based on $SO(10)$ SUSY GUTs~\cite{KM&Y} and $E_6$ SUSY GUTs~\cite{K&T}.
Sfermion sum rules at the TeV scale have been derived based on $SO(10)$ SUSY GUTs~\cite{Ch&H,A&P}.
We re-examine scalar sum rules at the TeV scale for various kinds of symmetry breaking patterns 
and particle assignments
by assuming that 1-loop RG effects and $D$-term contributions are dominant as a source to violate the scalar mass degeneracy.

~~\\
(1) $SU(5) \to G_{SM}$

{}From $SU(5)$ gauge symmetry, there exist following relations at the grand unification scale $M_U (= 2.1 \times 10^{16}$GeV),
\begin{eqnarray}
&~& {m}_{\tilde{q}_1}^{2}(M_U) = {m}_{\tilde{u}^*_R}^{2}(M_U) = {m}_{\tilde{e}^*_R}^{2}(M_U) \equiv m_{10}^{(1)2} ,
\label{m10-1} \\
&~& {m}_{\tilde{l}_{1}}^{2}(M_U) = {m}_{\tilde{d}^*_R}^{2}(M_U) \equiv {m}_{\overline{5}}^{(1)2} ,~ 
\label{m5-1} \\
&~& {m}_{\tilde{q}_2}^{2}(M_U) = {m}_{\tilde{c}^*_R}^{2}(M_U) = {m}_{\tilde{\mu}^*_R}^{2}(M_U) \equiv m_{10}^{(2)2} ,
\label{m10-2} \\
&~& {m}_{\tilde{l}_{2}}^{2}(M_U) = {m}_{\tilde{s}^*_R}^{2}(M_U) \equiv {m}_{\overline{5}}^{(2)2} ,~ 
\label{m5-2} \\
&~& {m}_{\tilde{q}_3}^{2}(M_U) = {m}_{\tilde{t}^*_R}^{2}(M_U) = {m}_{\tilde{\tau}^*_R}^{2}(M_U) \equiv m_{10}^{(3)2} ,
\label{m10-3} \\
&~& {m}_{\tilde{l}_{3}}^{2}(M_U) = {m}_{\tilde{b}^*_R}^{2}(M_U) \equiv {m}_{0}^{(3)2} ,~ 
\label{m5-3} \\
&~& m_{H_1}^2(M_U) \equiv m_{\overline{5}H}^2 ,~ m_{H_2}^2(M_U) \equiv m_{5H}^2 ,
\label{mH} 
\end{eqnarray}
where $m_{10}^{(1)}$, $m_{\overline{5}}^{(1)}$, $m_{10}^{(2)}$, $m_{\overline{5}}^{(2)}$, $m_{10}^{(3)}$, 
$m_{\overline{5}}^{(3)}$, $m_{\overline{5}H}$ and $m_{5H}$ are soft SUSY breaking scalar mass parameters.
Here we use the fact that $\tilde{q}_1$, $\tilde{u}^*_R$ and $\tilde{e}^*_R$ belong to $\bf{10}$, 
and $\tilde{l}_1$ and $\tilde{d}^*_R$ belong to $\overline{\bf{5}}$ and the same assignment holds on for second and third generations.
By eliminating those eight mass parameters, we obtain fifteen sum rules, the type I sfermion sum rules ((\ref{SR1}) and (\ref{SR2}))
and three kinds of sum rules ((\ref{SR3K}), (\ref{SR4K'}) and (\ref{SR5K}) with a common gaugino mass at $M_U$) such that,
\begin{eqnarray}
\hspace{-12mm}&~& M_{\tilde{u}_{L}}^{2}-M_{\tilde{u}_{R}}^{2} = M_{\tilde{c}_{L}}^{2}-M_{\tilde{c}_{R}}^{2} 
= M_{\tilde{t}_{L}}^{2}-M_{\tilde{t}_{R}}^{2} - F_t + F_b 
\nonumber \\
\hspace{-12mm}&~& ~~~ = \left(\zeta_2 \left(\frac{\alpha_2}{\alpha_3}\right)^2 - 15\zeta_1 \left(\frac{\alpha_1}{\alpha_3}\right)^2\right)
M_{\tilde{g}}^2
\nonumber \\
\hspace{-12mm}&~& ~~~~~~~~~~~~~~~~~~ +\left(\frac{4}{3}M_{W}^{2}-\frac{5}{6}M_{Z}^{2}\right)\cos 2\beta +5 \mathcal{S} , 
\label{SR3K-SU5} \\
\hspace{-12mm}&~& M_{\tilde{u}_{R}}^{2} - M_{\tilde{e}_{R}}^{2} = M_{\tilde{c}_{R}}^{2} - M_{\tilde{\mu}_{R}}^{2}  
\nonumber \\
\hspace{-12mm}&~& ~~~ = M_{\tilde{t}_{R}}^{2} - M_{\tilde{\tau}_{R}}^{2} - m_t^2 + 2 F_t -2F_{\tau}
\nonumber \\
\hspace{-12mm}&~& ~~~ = \left(\zeta_3 - 20\zeta_1 \left(\frac{\alpha_1}{\alpha_3}\right)^2\right)M_{\tilde{g}}^2
+\left(-\frac{5}{3}M_{W}^{2}+\frac{5}{3}M_{Z}^{2}\right)\cos 2\beta -10 \mathcal{S} .
\label{SR4K'-SU5} \\
\hspace{-12mm}&~& M_{\tilde{e}_{L}}^{2}-M_{\tilde{d}_{R}}^{2} = M_{\tilde{\mu}_{L}}^{2}-M_{\tilde{s}_{R}}^{2} 
= M_{\tilde{\tau}_{L}}^{2}-M_{\tilde{b}_{R}}^{2} + F_{\tau} - 2 F_b
\nonumber \\
\hspace{-12mm}&~& ~~~ = \left(- \zeta_3 + \zeta_2 \left(\frac{\alpha_2}{\alpha_3}\right)^2 + 5\zeta_1 \left(\frac{\alpha_1}{\alpha_3}\right)^2\right)
M_{\tilde{g}}^2 
\nonumber \\
\hspace{-12mm}&~& ~~~~~~~~~~~~~~~~~~ +\left(-\frac{4}{3}M_{W}^{2}+\frac{5}{6}M_Z^2\right)\cos 2\beta - 5 \mathcal{S} .
\label{SR5K-SU5}
\end{eqnarray}
The above sum rules (\ref{SR3K-SU5}) - (\ref{SR5K-SU5}) holds on in the case with the direct breakdown of
a grand unified symmetry to the SM one at $M_U$,
and hence the sfermion sector (and the grand unification) in SUSY GUT can be tested by using them.
We refer these sum rules (\ref{SR3K-SU5}) - (\ref{SR5K-SU5}) as the type IIA sfermion sum rules.

~~\\
(2) $SO(10) \to G_{SM}$

After the breakdown of $SO(10)$ gauge symmetry to the SM one, there exist following relations at the grand unification scale $M_U$,
\begin{eqnarray}
&~& {m}_{\tilde{q}_1}^{2}(M_U) = {m}_{\tilde{u}^*_R}^{2}(M_U) = {m}_{\tilde{e}^*_R}^{2}(M_U) = m_{16}^{(1)2} - D ,
\label{m16-1-D} \\
&~& {m}_{\tilde{l}_{1}}^{2}(M_U) = {m}_{\tilde{d}^*_R}^{2}(M_U) = m_{16}^{(1)2} + 3D ,
\label{m16-1+3D} \\
&~& {m}_{\tilde{q}_2}^{2}(M_U) = {m}_{\tilde{c}^*_R}^{2}(M_U) = {m}_{\tilde{\mu}^*_R}^{2}(M_U) = m_{16}^{(2)2} - D ,
\label{m16-2-D} \\
&~& {m}_{\tilde{l}_{2}}^{2}(M_U) = {m}_{\tilde{s}^*_R}^{2}(M_U) = m_{16}^{(2)2} +3D ,
\label{m16-2+3D} \\
&~& {m}_{\tilde{q}_3}^{2}(M_U) = {m}_{\tilde{t}^*_R}^{2}(M_U) = {m}_{\tilde{\tau}^*_R}^{2}(M_U) = m_{16}^{(3)2} -D ,
\label{m16-3-D} \\
&~& {m}_{\tilde{l}_{3}}^{2}(M_U) = {m}_{\tilde{b}^*_R}^{2}(M_U) = m_{16}^{(3)2} + 3D ,
\label{m16-3+3D} \\
&~& m_{H_1}^2(M_U) = m_{10}^2 - 2D ,~ m_{H_2}^2(M_U) = m_{10}^2 + 2D ,
\label{mH} 
\end{eqnarray}
where $m_{16}^{(1)}$, $m_{16}^{(2)}$, $m_{16}^{(3)}$ and $m_{10}$ are soft SUSY breaking scalar mass parameters
and $D$ is a parameter which represents $D$-term condensation related to $SO(10)/SU(5)$ generator.
Here we assume that the particle assignment is the ordinary one 
where particles in each generation belong to $\bf{16}$
and two Higgs doublets belong to $\bf{10}$.
By eliminating those five unknown parameters, we obtain eighteen sum rules, i.e., fifteen ones
 (\ref{SR1}), (\ref{SR2}), (\ref{SR3K-SU5}), (\ref{SR4K'-SU5}) and (\ref{SR5K-SU5}), which are same as those derived from $SU(5)$ breaking, 
and the following three sum rules,
\begin{eqnarray}
&~& m^{2}_{H_{1}} - m^{2}_{H_{2}} - M_{\tilde{u}_{R}}^{2} + M_{\tilde{d}_{R}}^{2} 
\nonumber\\
&~& ~~~ = -12\zeta_1 M_1^2+\left(M_{W}^{2}-M_{Z}^{2}\right)\cos 2\beta - F_{\tau} - 3F_{b} + 3F_{t} ,
\label{SO10-1}\\
&~& M_{\tilde{u}_{R}}^{2} - M_{\tilde{d}_{R}}^{2} = M_{\tilde{c}_{R}}^{2} - M_{\tilde{s}_{R}}^{2} 
= M_{\tilde{t}_{R}}^{2} - M_{\tilde{b}_{R}}^{2} -m_{t}^2 +2F_{t} -2F_{b} .
\label{SO10-2}
\end{eqnarray}

In Ref.\cite{IK&Y}, scalar sum rules are examined in the case 
that some low-energy particles appear as a linear combination of several distinct fields.
We refer $\lq$particle twisting' as such a linear combination of some particles.
In the case with particle twisting concerning ($\tilde{l}_1$, $\tilde{d}^*_R$), ($\tilde{l}_2$, $\tilde{s}^*_R$) 
and ($\tilde{l}_3$, $\tilde{b}^*_R$), i.e., each one constitutes as a combination of particles 
in $\bf{16}$ and those in $\bf{10}$,
the sum rules (\ref{SO10-1}) and (\ref{SO10-2}) do not hold on.

~~\\
(3) $SU(5) \times U(1)_{F} \to G_{SM}$

The particle assigment is different from the ordinary $SU(5)$ GUT, and
the model is called $\lq$flipped $SU(5)$ GUT'.
In this model, $\tilde{q}_1$, $\tilde{d}^*_R$ and $\tilde{\nu}^*_R$ belong to $\bf{10}$, $\tilde{l}_1$ and $\tilde{u}^*_R$ belong to $\overline{\bf{5}}$ 
and $\tilde{e}^*_R$ belongs to $\bf{1}$ in $SU(5)$.
After the breakdown of $SU(5) \times U(1)_{F}$ gauge symmetry to the SM one, 
there exist following relations among the first generation sfermions at the breaking scale $M_U$,
\begin{eqnarray}
&~& {m}_{\tilde{q}_1}^{2}(M_U) = m_{10}^{(1)2} - (4g_5^2 + g_F^2) D_F ,
\label{F-1} \\
&~& {m}_{\tilde{d}^*_R}^{2}(M_U)  = m_{10}^{(1)2} + (16g_5^2 - g_F^2)D_F ,
\label{F-2} \\
&~& {m}_{\tilde{u}^*_R}^{2}(M_U) = m_{\overline{5}}^{(1)2} + (-8g_5^2 + 3g_F^2)D_F ,
\label{F-3} \\
&~& {m}_{\tilde{l}_1}^{2}(M_U)  = m_{\overline{5}}^{(1)2} + (12g_5^2 + 3 g_F^2) D_F ,
\label{F-4} \\
&~& {m}_{\tilde{e}^*_R}^{2}(M_U)  = m_{1}^{(1)2} - 5g_F^2 D_F ,
\label{F-5} 
\end{eqnarray}
where $m_{10}^{(1)}$, $m_{\overline{5}}^{(1)}$ and $m_{1}^{(1)}$ are soft SUSY breaking scalar mass parameters
and $D_F$ is a parameter which represents the $D$-term condensation related to $SU(5) \times U(1)_F/G_{SM}$ generator.
The $g_5$ and $g_F$ are gauge coupling constants of $SU(5)$ and $U(1)_F$, respectively.
The same type of relations hold on for second and third generations and Higgs bosons.
In this case, there are twelve arbitrary parameters, i.e., eleven soft SUSY breaking scalar mass parameters
and one parameter from the $D$-term contribution.
By eliminating these parameters, we obtain eleven sum rules, i.e., the type I sfermion sum rules and five ones such that
\begin{eqnarray}
&~& M_{\tilde{u}_{L}}^{2} + M^{2}_{\tilde{d}_{L}} -2M_{\tilde{u}_{R}}^{2} -2M^{2}_{\tilde{d}_{R}}
+M_{\tilde{\nu}_{eL}}^{2} + M^{2}_{\tilde{e}_{L}} 
\nonumber \\
&~& ~~~ = M_{\tilde{c}_{L}}^{2} + M^{2}_{\tilde{s}_{L}} -2M_{\tilde{c}_{R}}^{2} -2M^{2}_{\tilde{s}_{R}}
+M_{\tilde{\nu}_{\mu L}}^{2} + M^{2}_{\tilde{\mu}_{L}} 
\nonumber \\
&~& ~~~ = M_{\tilde{t}_{L}}^{2} + M^{2}_{\tilde{b}_{L}} -2M_{\tilde{t}_{R}}^{2} -2M^{2}_{\tilde{b}_{R}}
+M_{\tilde{\nu}_{\tau L}}^{2} + M^{2}_{\tilde{\tau}_{L}} + m_t^2 
\nonumber \\
&~& ~~~~~~~  - 2 F_t - 2F_b + 2F_{\tau}
\nonumber \\
&~& ~~~ = -2 \left(\zeta_3 - 2 \zeta_2 \left(\frac{\alpha_2}{\alpha_3}\right)^2\right)
M_{\tilde{g}}^2 -20\zeta_1 M_1^2 ,
\label{SU5F-1}\\
&~& M_{\tilde{d}_{L}}^{2} - M^{2}_{\tilde{d}_{R}} = M_{\tilde{s}_{L}}^{2} - M^{2}_{\tilde{s}_{R}} 
= M_{\tilde{b}_{L}}^{2} - M^{2}_{\tilde{b}_{R}} + F_t - F_b ,
\label{SU5F-2}
\end{eqnarray}
where we use the ralation $M_{\tilde{g}} = \alpha_3 M_2/\alpha_2$
due to the partial unification of gaugino masses, i.e., $M_3(M_U) = M_2(M_U)$.
Note that $M_1(M_U)$ does not necessarily equal to $M_3(M_U)$.

~~\\
(4) $SU(4)\times SU(2)_{L}\times SU(2)_{R} \to G_{SM}$

In this case, there are eight arbitrary parameters, i.e., seven soft SUSY breaking scalar mass parameters
and one parameter from a $D$-term contribution.
By eliminating them, we obtain twelve sum rules, i.e., the type I sfermion sum rules and nine ones such that
\begin{eqnarray}
&~& M_{\tilde{u}_{L}}^{2} + M_{\tilde{d}_{L}}^{2}-M_{\tilde{\nu}_{eL}}^{2}-M_{\tilde{e}_{L}}^{2} -2M_{\tilde{e}_{R}}^{2} +2M_{\tilde{d}_{R}}^{2}
\nonumber \\
&~& ~~~ = M_{\tilde{c}_{L}}^{2} + M_{\tilde{s}_{L}}^{2}-M_{\tilde{\nu}_{\mu L}}^{2}-M_{\tilde{\mu}_{L}}^{2} 
 -2M_{\tilde{\mu}_{R}}^{2} +2M_{\tilde{s}_{R}}^{2}
\nonumber \\
&~& ~~~ = M_{\tilde{t}_{L}}^{2} + M_{\tilde{b}_{L}}^{2}-M_{\tilde{\nu}_{\tau L}}^{2}-M_{\tilde{\tau}_{L}}^{2} 
 -2M_{\tilde{\tau}_{R}}^{2} +2M_{\tilde{b}_{R}}^{2} - m_t^2 
\nonumber \\
&~& ~~~~~~~  + 2 F_t + 6F_b -6F_{\tau}
\nonumber \\
&~& ~~~ = 2\zeta_3 M_3^2 -40\zeta_1 M_1^2 ,
\label{PS-1} \\
&~& g_{R}^{2}\left(M_{\tilde{u}_{L}}^{2}+M_{\tilde{d}_{L}}^{2}-M_{\tilde{\nu}_{eL}}^{2}-M_{\tilde{e}_L}^{2}\right)
 -2g_{4}^{2}\left(M_{\tilde{u}_{R}}^{2}-M_{\tilde{d}_{R}}^{2}\right)
\nonumber \\
&~& ~~~ = g_{R}^{2}\left(M_{\tilde{c}_{L}}^{2}+M_{\tilde{s}_{L}}^{2}-M_{\tilde{\nu}_{\mu L}}^{2}-M_{\tilde{\mu}_L}^{2}\right)
 -2g_{4}^{2}\left(M_{\tilde{c}_{R}}^{2}-M_{\tilde{s}_{R}}^{2}\right)
\nonumber \\
&~& ~~~ = g_{R}^{2}\left(M_{\tilde{t}_{L}}^{2}+M_{\tilde{b}_{L}}^{2}-M_{\tilde{\nu}_{\tau L}}^{2}-M_{\tilde{\tau}_L}^{2}\right)
 -2g_{4}^{2}\left(M_{\tilde{t}_{R}}^{2}-M_{\tilde{b}_{R}}^{2}\right) 
\nonumber \\
&~& ~~~~~~ - g_R^2\left(m_t^2 - 2 F_t - 2F_b + 2F_{\tau}\right)
 + 2 g_4^2\left(m_t^2 - 2 F_t + 2F_b\right)
\nonumber \\
&~& ~~~ = 2g_{R}^{2}\zeta_3 M_3^2-\left(16g_{R}^{2}+24g_{4}^{2}\right)\zeta_1 M_1^2
\nonumber\\
&~& ~~~~~~ + \left(\frac{4}{3}g_{R}^{2}-2g_{4}^{2}\right)\left(M_{W}^{2}-M_{Z}^{2}\right)\cos 2\beta +\left(8g_{R}^{2}+12g_{4}^{2}\right) \mathcal{S} ,
\label{PS-2} \\
&~& m_{H_{1}}^{2}-m_{H_{2}}^{2}-M_{\tilde{u}_{R}}^{2}-M_{\tilde{d}_{R}}^{2}
=m_{H_{1}}^{2}-m_{H_{2}}^{2}-M_{\tilde{c}_{R}}^{2}-M_{\tilde{s}_{R}}^{2}
\nonumber\\
&~& ~~~ =m_{H_{1}}^{2}-m_{H_{2}}^{2}-M_{\tilde{t}_{R}}^{2}-M_{\tilde{b}_{R}}^{2} + m_t^2 - 2 F_t - 2F_b
\nonumber\\
&~& ~~~ = -12\zeta_1 M_1^2+\left(M_{W}^{2}-M_{Z}^{2}\right)\cos 2\beta +3F_{t}-3F_{b}-F_{\tau} ,
\label{PS-3}
\end{eqnarray}
where we assume that $H_1$ and $H_2$ belong to $({\bf 1}, {\bf 2}, {\bf 2})$ of $SU(4)\times SU(2)_{L}\times SU(2)_{R}$.

~~\\
(5) $SU(3)_{c}\times SU(2)_{L}\times SU(2)_{R}\times U(1)_{B-L} \to G_{SM}$

In this case, there are fourteen arbitrary parameters, i.e., thirteen soft SUSY breaking scalar mass parameters
and one parameter from a $D$-term contribution.
By eliminating these parameters, we obtain nine sum rules (\ref{SR1}), (\ref{SR2}) and (\ref{PS-3}).

~~\\
(6) $E_6\to G_{SM}$

In the case without particle twisting, there are two kinds of particle assignment in $SO(10)$ subgroup.
One is that $\tilde{l}_1$ and $\tilde{d}^*_R$ belong to $\bf{16}$ and $H_1$ belongs to $\bf{10}$.
The other is that $\tilde{l}_1$ and $\tilde{d}^*_R$ belong to $\bf{10}$ and $H_1$ belongs to $\bf{16}$.
For both ones, we can obtain the same eighteen sum rules as those with $SO(10) \to G_{SM}$.

In the case with particle twisting concerning ($\tilde{l}_1$, $\tilde{d}^*_R$), ($\tilde{l}_2$, $\tilde{s}^*_R$) 
and ($\tilde{l}_3$, $\tilde{b}^*_R$), sum rules are reduced to fifteen ones,
the type I ((\ref{SR1}) and (\ref{SR2})) and the type IIA ((\ref{SR3K-SU5}), (\ref{SR4K'-SU5}) and (\ref{SR5K-SU5})).
As pointed out in \cite{IK&Y}, the twisting angles are determined from sparticle spectrum,
and hence the flavor structure of GUT can be probed.

~~\\
(7) $SU(6)\times SU(2)_{R} \to G_{SM}$

For sfermions in the first generation, $\tilde{q}_1$ and $\tilde{l}_1$ belong to $({\bf 15}, {\bf 1})$ and
$\tilde{u}^*_R$, $\tilde{d}^*_R$ and $\tilde{e}^*_R$ belong to $(\overline{\bf 6}, {\bf 2})$ of $SU(6)\times SU(2)_{R}$.
The same assignment holds on for other generations.
In this case, we can obtain the same sum rules as those with $SU(4)\times SU(2)_{L}\times SU(2)_{R} \to G_{SM}$
by replacing $g_4$ by the gauge coupling of $SU(6)$.

~~\\
(8) $SU(6)\times SU(2)_{L} \to G_{SM}$

For sfermions in the first generation, $\tilde{q}_1$ and $\tilde{l}_1$ belong to $({\bf 6}, {\bf 2})$ and
$\tilde{u}^*_R$, $\tilde{d}^*_R$ and $\tilde{e}^*_R$ belong to $(\overline{\bf 15}, {\bf 1})$ of $SU(6)\times SU(2)_{L}$.
The same assignment holds on for other generations.
In this case, there are eight arbitrary parameters, i.e., six soft SUSY breaking scalar mass parameters
and two parameters from $D$-term contributions.
By eliminating these parameters, we obtain fifteen sum rules, i.e., twelve relations (\ref{SR1}), (\ref{SR2}), (\ref{PS-1}), and (\ref{PS-3}),
and three ones such that
\begin{eqnarray}
&~& M_{\tilde{u}_{R}}^{2} - M_{\tilde{e}_{R}}^{2} = M_{\tilde{c}_{R}}^{2} - M_{\tilde{\mu}_{R}}^{2}  
\nonumber \\
&~& ~~~ = M_{\tilde{t}_{R}}^{2} - M_{\tilde{\tau}_{R}}^{2} - m_t^2 + 2 F_t -2F_{\tau}
\nonumber \\
&~& ~~~ = \zeta_3 M_3^2-20\zeta_1 M_1^2+\left(-\frac{5}{3}M_{W}^{2}+\frac{5}{3}M_{Z}^{2}\right)\cos 2\beta -10 \mathcal{S} .
\label{SU6} 
\end{eqnarray}

~~\\
(9) $SU(3)_{c}\times SU(3)_{L}\times SU(3)_{R} \to G_{SM}$

For sfermions in the first generation, $\tilde{q}_1$ belongs to $({\bf 3}, {\bf 3}, {\bf 1})$,
$\tilde{u}^*_R$ and $\tilde{d}^*_R$ belong to $(\overline{\bf 3}, {\bf 1}, \overline{\bf 3})$ 
and $\tilde{l}_L$ and $\tilde{e}^*_R$ belong to $({\bf 1}, \overline{\bf 3}, {\bf 3})$ of $SU(6)\times SU(2)_{R}$.
The same assignment holds on for other generations.
In this case, there are twelve arbitrary parameters, i.e., ten soft SUSY breaking scalar mass parameters
and two parameters from $D$-term contributions.
By eliminating these parameters, we obtain eleven sum rules, i.e., nine relations (\ref{SR1}), (\ref{SR2}) and (\ref{PS-3}),
and two ones such that
\begin{eqnarray}
M_{\tilde{e}_L}^{2} - M_{\tilde{e}_R}^{2} = M_{\tilde{\mu}_L}^{2} - M_{\tilde{\mu}_R}^{2} 
= M_{\tilde{\tau}_L}^{2} - M_{\tilde{\tau}_R}^{2} - F_{\tau} .
\label{SU3-1}
\end{eqnarray}

\subsection{Orbifold grand unification in five dimension}

Higher-dimensional SUSY GUTs  on an orbifold have attractive features as a realistic model beyond the MSSM.
The triplet-doublet splitting of Higgs multiplets is elegantly realized in SUSY $SU(5)$ GUT
in a five dimension.\cite{K}\cite{H&N}
Four-dimensional chiral fermions are generated through the dimensional reduction.
Those phenomena originates from the fact that a part of zero modes are projected out by orbifolding, i.e., by
non-trivial boundary conditions (BCs) concerning extra dimensions on bulk fields.
Therefore we expect that some specific sum rules obtained in the previous section can be modified
by splitting a bulk multiplet on the orbifold breaking.
Chiral anomalies may arise at the boundaries with the advent of chiral fermions.
Those anomalies must be cancelled in the four-dimensional effective theory
by the contribution of brane chiral fermions and/or counter terms such as the
Chern-Simons term.\cite{C&H,KK&L} 

Recently, the possibility of complete family unification has been studied in SUSY $SU(N)$ GUTs 
defined on a five-dimensional space-time $M^4 \times (S^1/Z_2)$.\cite{KK&O}
Here $M^4$ is the four-dimensional Minkowski space-time 
and $S^1/Z_2$ is the one-dimensional orbifold. Interesting models have been found, 
in which bulk fields from a single hypermultiplet and a few brane fields compose three families.

In this section, we derive peculiar sum rules among sfermion masses in two kinds of models where
a partial family unification is realized, after studying $SU(5)$ SUSY orbifold GUT as a warm-up.\footnote{
Sfermion masses have been studied from the viewpoint of flavor symmetry and its violation
in $SU(5)$ SUSY orbifold GUT.\cite{H&N2}}
We assume that extra gauge symmetries are broken at the same time as the orbifold breaking,
and we do not specify the origin of soft SUSY breaking terms though some SUSY breaking mechanisms
are known on the orbifold.\footnote{The typical one is Scherk-Schwarz mechanism, in which SUSY
is broken by the difference of BCs between bosons and fermions.\cite{S&S}
This mechanism on $S^1/Z_2$ leads to a restricted type of soft SUSY breaking parameters such as
$M_i = \beta/R$ for bulk gauginos and $m_{\tilde{F}}^2 = (\beta/R)^2$ for bulk scalar particles
where $\beta$ is a real parameter and $R$ is a radius of $S^1$.}
We also assume that the four-dimensional effective theory is anomaly free due to the presence
of appropriate brane fields and/or the Chern-Simons term.

~~\\
(a) $SU(5) \to G_{SM}$

The orbifold breaking $SU(5) \to G_{SM}$ is realized by the $Z_2$ parity assignment,
\begin{eqnarray}
&~& P_0 = {\rm diag}(+1, +1, +1, +1, +1) ~, 
\label{P0-SU5} \\
&~& P_1 = {\rm diag}(+1, +1, +1, -1, -1) ~.
\label{P1-SU5}
\end{eqnarray}
After the breakdown of $SU(5)$, $\overline{\bf{5}}$ and $\bf{10}$ are decomposed into a sum of multiplets 
regarding to the maximal subgroup $SU(3)_C \times SU(2)_L \times U(1)_Y$, 
\begin{eqnarray}
&~& \overline{\bf{5}} = (\overline{\bf{3}}, {\bf{1}}, 1/3) + ({\bf{1}}, {\bf{2}}, -1/2) ,
\label{bar5}\\
&~& {\bf{10}} = ({\bf{3}}, {\bf{2}}, 1/6) + (\overline{\bf{3}}, {\bf{1}}, -2/3) + ({\bf{1}}, {\bf{1}}, 1) ,
\label{10}
\end{eqnarray}
respectively.
The $Z_2$ parity ($\mathcal{P}_0, \mathcal{P}_1$) of multiplets is listed in Table \ref{T0}.
In the first column, the subscript $L$ ($R$) represents the left-handedness (right-handedness) for Weyl fermions.
In the second and third column, ($\eta_{\overline{5}}$, $\eta_{10}$) and ($\eta'_{\overline{5}}$, $\eta'_{10}$) are intrinsic $Z_2$ parities 
which take a value $+1$ or $-1$.

\begin{table}
\caption{$Z_2$ parity ($\mathcal{P}_0, \mathcal{P}_1$) for Weyl fermions from $\overline{\bf{5}}$ and ${\bf{10}}$}
\label{T0}
\begin{center}
\begin{tabular}{c|c|c} \hline
{\it Representation} & $\mathcal{P}_0$ & $\mathcal{P}_1$   \\ \hline\hline
$(\overline{\bf{3}}, {\bf{1}}, 1/3)_L$ &  $\eta_{\overline{5}}$ & $\eta'_{\overline{5}}$ \\
$(\overline{\bf{3}}, {\bf{1}}, 1/3)_R$ &  $-\eta_{\overline{5}}$ & $-\eta'_{\overline{5}}$ \\
$({\bf{1}}, {\bf{2}}, -1/2)_L$ &  $\eta_{\overline{5}}$ & $-\eta'_{\overline{5}}$ \\
$({\bf{1}}, {\bf{2}}, -1/2)_R$ &  $-\eta_{\overline{5}}$ & $\eta'_{\overline{5}}$ \\ \hline
$({\bf{3}}, {\bf{2}}, 1/6)_L$ &  $\eta_{10}$ & $-\eta'_{10}$ \\
$({\bf{3}}, {\bf{2}}, 1/6)_R$ &  $-\eta_{10}$ & $\eta'_{10}$ \\
$(\overline{\bf{3}}, {\bf{1}}, -2/3)_L$ & $\eta_{10}$ & $\eta'_{10}$ \\
$(\overline{\bf{3}}, {\bf{1}}, -2/3)_R$ & $-\eta_{10}$ & $-\eta'_{10}$ \\
$({\bf{1}}, {\bf{1}}, 1)_L$ & $\eta_{10}$ & $\eta'_{10}$ \\
$({\bf{1}}, {\bf{1}}, 1)_R$ & $-\eta_{10}$ & $-\eta'_{10}$ \\ \hline
\end{tabular}
\end{center}
\end{table}

Whatever intrinsic $Z_2$ parity for $\overline{\bf{5}}$ we choose, either zero mode of 
$(\overline{\bf{3}}, {\bf{1}}, 1/3)$ or $({\bf{1}}, {\bf{2}}, -1/2)$ is projected out and hence we, in general, have no scalar mass relation 
from a bulk field whose representation is $\overline{\bf{5}}$.
For a bulk fields with ${\bf{10}}$, $(\overline{\bf{3}}, {\bf{1}}, -2/3)_L$ and $({\bf{1}}, {\bf{1}}, 1)_L$ have zero mode
when we take $\eta_{10} = 1$ and $\eta'_{10} = 1$.
In this case, there exists the following relation at a breaking scale $M_U$,
\begin{eqnarray}
{m}_{\tilde{u}^*_{R}}^{2}(M_U) = {m}_{\tilde{e}^*_{R}}^{2}(M_U) ,
\label{5D-SU(5)}
\end{eqnarray}
where zero mode of superpartners for $(\overline{\bf{3}}, {\bf{1}}, -2/3)_L$ and $({\bf{1}}, {\bf{1}}, 1)_L$ are identified with
$\tilde{u}^*_R$ and $\tilde{e}^*_R$, respectively.
The relation (\ref{5D-SU(5)}) leads to the sum rule (a piece of type IIA sfermion sum rules),
\begin{eqnarray}
M_{\tilde{u}_{R}}^{2} - M_{\tilde{e}_{R}}^{2} 
 = \left(\zeta_3 -20\zeta_1\left(\frac{\alpha_1}{\alpha_3}\right)^2\right)M_{\tilde{g}}^2
 -\frac{5}{3} \left(M_{W}^{2} - M_{Z}^{2}\right)\cos 2\beta -10 \mathcal{S} ,
\label{typeIIB} 
\end{eqnarray}
where we use the GUT relation $M_3(M_U) = M_2(M_U)$.
This type of sum rule generally holds on orbifold breaking of $SU(N)$ gauge symmetry for bulk fields with 
an antisymmetric representation if the bulk field contains ${\bf 10}_L$ or ${\overline{\bf 10}}_R$ under the subgroup $SU(5)$,
and those $SU(2)_L$ singlets have even $Z_2$ parities.
We refer a piece of type IIA such as (\ref{typeIIB}) as the type IIB sfermion sum rules.

~~\\
(b) $SU(8) \to G_{SM} \times SU(3) \times U(1)'$

We study sfermion mass relations stemed from $[8, 4]$ of $SU(8)$ after the orbifold breaking 
$SU(8) \to G_{SM} \times SU(3) \times U(1)'$, which is realized by the $Z_2$ parity assignment,
\begin{eqnarray}
&~& P_0 = {\rm diag}(+1, +1, +1, +1, +1, -1, -1, -1) ~, 
\label{P0-SU8} \\
&~& P_1 = {\rm diag}(+1, +1, +1, -1, -1, +1, +1, +1) ~.
\label{P1-SU8}
\end{eqnarray}
After the breakdown of $SU(8)$, the fourth antisymmetric representation $[8, 4]$ with ${}_{8}C_{4}$ components
are decomposed into a sum of multiplets regarding to the subgroup $SU(3)_C \times SU(2)_L \times SU(3)$, 
\begin{eqnarray}
[8, 4] = \sum_{l_1 =0}^{4} \sum_{l_2 = 0}^{4-l_1} \left({}_{3}C_{l_1}, {}_{2}C_{l_2}, {}_{3}C_{4-l_1-l_2}\right) ,
\label{84}
\end{eqnarray}
where $l_1$ and $l_2$ are intergers.
The $Z_2$ parity of $\left({}_{3}C_{l_1}, {}_{2}C_{l_2}, {}_{3}C_{4-l_1-l_2}\right)$ is given by
\begin{eqnarray}
\mathcal{P}_0 =  (-1)^{l_1+l_2} \eta_{[8,4]} ,~ P_1 = (-1)^{l_2} \eta'_{[8,4]} ,
\label{Z2-SU8}
\end{eqnarray}
where $\eta_{[8,4]}$ and $\eta'_{[8,4]}$ are intrinsic $Z_2$ parities which take a value $+1$ or $-1$.
We assume that the $Z_2$ parity (\ref{Z2-SU8}) is assigned for the left-handed Weyl fermions.
The corresponding right-handed ones have opposite $Z_2$ parities.
In Table \ref{T1}, the $Z_2$ parity ($\mathcal{P}_0, \mathcal{P}_1$) and the $U(1)'$ charge are listed for representations of Weyl fermions.
In the first column, the subscript $L$ ($R$) represents the left-handedness (right-handedness) for Weyl fermions.

\begin{table}
\caption{$Z_2$ parity ($\mathcal{P}_0, \mathcal{P}_1$) and the $U(1)'$ charge for Weyl fermions from $[8,4]$}
\label{T1}
\begin{center}
\begin{tabular}{c|c|c|c} \hline
{\it Representation} & $\mathcal{P}_0$ & $\mathcal{P}_1$ & $U(1)'$  \\ \hline\hline
$\left({}_{3}C_1, {}_{2}C_{0}, {}_{3}C_3\right)_L$ & $-\eta_{[8,4]}$ & $\eta'_{[8,4]}$ & $-12$ \\
$\left({}_{3}C_1, {}_{2}C_{0}, {}_{3}C_3\right)_R$ & $\eta_{[8,4]}$ & $- \eta'_{[8,4]}$ & $-12$ \\
$\left({}_{3}C_{0}, {}_{2}C_1, {}_{3}C_{3}\right)_L$ & $-\eta_{[8,4]}$ & $-\eta'_{[8,4]}$ & $-12$ \\ 
$\left({}_{3}C_{0}, {}_{2}C_1, {}_{3}C_{3}\right)_R$ & $\eta_{[8,4]}$ & $\eta'_{[8,4]}$ & $-12$ \\ \hline
$\left({}_{3}C_{2}, {}_{2}C_{0}, {}_{3}C_{2}\right)_L$ & $\eta_{[8,4]}$ & $\eta'_{[8,4]}$ & $-4$ \\
$\left({}_{3}C_{2}, {}_{2}C_{0}, {}_{3}C_{2}\right)_R$ & $-\eta_{[8,4]}$ & $- \eta'_{[8,4]}$ & $-4$ \\
$\left({}_{3}C_{0}, {}_{2}C_{2}, {}_{3}C_{2}\right)_L$ & $\eta_{[8,4]}$ & $\eta'_{[8,4]}$ & $-4$ \\
$\left({}_{3}C_{0}, {}_{2}C_{2}, {}_{3}C_{2}\right)_R$ & $-\eta_{[8,4]}$ & $- \eta'_{[8,4]}$ & $-4$ \\
$\left({}_{3}C_{1}, {}_{2}C_{1}, {}_{3}C_{2}\right)_L$ & $\eta_{[8,4]}$ & $-\eta'_{[8,4]}$ & $-4$ \\ 
$\left({}_{3}C_{1}, {}_{2}C_{1}, {}_{3}C_{2}\right)_R$ & $-\eta_{[8,4]}$ & $\eta'_{[8,4]}$ & $-4$ \\ \hline
$\left({}_{3}C_{3}, {}_{2}C_{0}, {}_{3}C_{1}\right)_L$ & $-\eta_{[8,4]}$ & $\eta'_{[8,4]}$ & $4$ \\
$\left({}_{3}C_{3}, {}_{2}C_{0}, {}_{3}C_{1}\right)_R$ & $\eta_{[8,4]}$ & $-\eta'_{[8,4]}$ & $4$ \\
$\left({}_{3}C_{1}, {}_{2}C_{2}, {}_{3}C_{1}\right)_L$ & $-\eta_{[8,4]}$ & $\eta'_{[8,4]}$ & $4$ \\
$\left({}_{3}C_{1}, {}_{2}C_{2}, {}_{3}C_{1}\right)_R$ & $\eta_{[8,4]}$ & $-\eta'_{[8,4]}$ & $4$ \\
$\left({}_{3}C_{2}, {}_{2}C_{1}, {}_{3}C_{1}\right)_L$ & $-\eta_{[8,4]}$ & $-\eta'_{[8,4]}$ & $4$ \\ 
$\left({}_{3}C_{2}, {}_{2}C_{1}, {}_{3}C_{1}\right)_R$ & $\eta_{[8,4]}$ & $\eta'_{[8,4]}$ & $4$ \\ \hline
$\left({}_{3}C_{3}, {}_{2}C_{1}, {}_{3}C_{0}\right)_L$ & $\eta_{[8,4]}$ & $-\eta'_{[8,4]}$ & $12$ \\
$\left({}_{3}C_{3}, {}_{2}C_{1}, {}_{3}C_{0}\right)_R$ & $-\eta_{[8,4]}$ & $\eta'_{[8,4]}$ & $12$ \\
$\left({}_{3}C_{2}, {}_{2}C_{2}, {}_{3}C_{0}\right)_L$ & $\eta_{[8,4]}$ & $\eta'_{[8,4]}$ & $12$ \\ 
$\left({}_{3}C_{2}, {}_{2}C_{2}, {}_{3}C_{0}\right)_R$ & $-\eta_{[8,4]}$ & $-\eta'_{[8,4]}$ & $12$ \\ \hline
\end{tabular}
\end{center}
\end{table}

Let us take $\eta_{[8,4]} = 1$ and $\eta'_{[8,4]} =1$.
In this case, particles with even $Z_2$ parities are given in Table \ref{T2}.
Each particle possesses a zero mode whose scalar component is identified with one of the MSSM particles in four dimension.

\begin{table}
\caption{Sfermions with even $Z_2$ parity from $[8, 4]$}
\label{T2}
\begin{center}
\begin{tabular}{c|c|c} \hline
{\it Representation} & {\it Species} & $U(1)'$ \\ \hline
$\left(\overline{\bf{3}}, {\bf{1}}, {\bf{1}}\right)$ & $\tilde{d}^*_R$ (or $\tilde{s}^*_R$, $\tilde{b}^*_R$) & 12 \\
$\left({\bf{3}}, {\bf{2}}, \overline{\bf{3}}\right)$ & $\tilde{q}_{1}$, $\tilde{q}_{2}$, $\tilde{q}_{3}$  & $-4$ \\
$\left({\bf{1}}, {\bf{1}}, \overline{\bf{3}}\right)$ & $\tilde{e}^*_R$, $\tilde{\mu}^*_R$, $\tilde{\tau}^*_R$ & $-4$ \\
$\left(\overline{\bf{3}}, {\bf{1}}, \overline{\bf{3}}\right)$ & $\tilde{u}^*_R$, $\tilde{c}^*_R$, $\tilde{t}^*_R$ & $-4$ \\
$\left({\bf{1}}, {\bf{2}}, {\bf{1}}\right)$ & $\tilde{l}_{1}$ (or $\tilde{l}_{2}$, $\tilde{l}_{3}$) & $12$ \\ \hline
\end{tabular}
\end{center}
\end{table}

After the breakdown of $SU(3) \times U(1)'$ gauge symmetry, 
there exist the following relations at a breaking scale $M_U$,
\begin{eqnarray}
&~& {m}_{\tilde{d}^*_R}^{2}(M_U) = m_{[8,4]}^{2} + 12 D' ,
\label{SU(8)-d}\\
&~& {m}_{\tilde{q}_{1}}^{2}(M_U) = m_{[8,4]}^{2} - D_1 - D_2 - 4 D' ,
\label{SU(8)-q1}\\
&~& {m}_{\tilde{q}_{2}}^{2}(M_U) = m_{[8,4]}^{2} + D_1 - D_2 - 4 D' ,
\label{SU(8)-q2}\\
&~& {m}_{\tilde{q}_{3}}^{2}(M_U) = m_{[8,4]}^{2}  + 2 D_2 - 4 D' ,
\label{SU(8)-q3}\\
&~& {m}_{\tilde{u}^*_{R}}^{2}(M_U) = {m}_{\tilde{e}^*_{R}}^{2}(M_U) = m_{[8,4]}^{2} - D_1 - D_2 - 4 D' ,
\label{SU(8)-u}\\
&~& {m}_{\tilde{c}^*_{R}}^{2}(M_U) = {m}_{\tilde{\mu}^*_{R}}^{2}(M_U) = m_{[8,4]}^{2} + D_1 - D_2 - 4 D' ,
\label{SU(8)-c}\\
&~& {m}_{\tilde{t}^*_{R}}^{2}(M_U) = {m}_{\tilde{\tau}^*_{R}}^{2}(M_U) = m_{[8,4]}^{2}  + 2 D_2 - 4 D' ,
\label{SU(8)-t}\\
&~& {m}_{\tilde{l}_{1}}^{2}(M_U) = m_{[8,4]}^{2} + 12 D' ,
\label{SU(8)-l1}
\end{eqnarray}
where $m_{[8,4]}$ is a soft SUSY breaking scalar mass parameter, $D_1$ and $D_2$ are parameters which represent $D$-term condensations 
related to $SU(3)$ generator
and $D'$ stands for the $D$-term contribution of $U(1)'$.
By eliminating those four unknown parameters, we obtain the following relations at $M_U$,
\begin{eqnarray}
&~& {m}_{\tilde{u}^*_{R}}^{2}(M_U) = {m}_{\tilde{e}^*_{R}}^{2}(M_U) = {m}_{\tilde{q}_{1}}^{2}(M_U) ,~ 
\label{SU(8)-1} \\
&~& {m}_{\tilde{c}^*_{R}}^{2}(M_U) = {m}_{\tilde{\mu}^*_{R}}^{2}(M_U) = {m}_{\tilde{q}_{2}}^{2}(M_U) ,~
\label{SU(8)-2} \\
&~& {m}_{\tilde{t}^*_{R}}^{2}(M_U) = {m}_{\tilde{\tau}^*_{R}}^{2}(M_U) = {m}_{\tilde{q}_{3}}^{2}(M_U) ,
\label{SU(8)-3}\\
&~& {m}_{\tilde{d}^*_R}^{2}(M_U) = {m}_{\tilde{l}_{1}}^{2}(M_U) .
\label{SU(8)-4}
\end{eqnarray}
We need two right-handed down-type quark chiral supermultiplets and two $SU(2)$ lepton doublet
chiral supermultiplets in order to obtain three generations.
We assume that they are brane fields.
When our four-dimensional world is a boundary on the point $y = 0$, each chiral multiplet must form a multiplet of $SU(5)$.
Hence we have complete $SU(5)$ type of sum rules in the case that three generations come from $[8, 4]$ 
and two brane fields with $\overline{5}$.
This result occurs from the fact that $[8, 4]$ is a real representation.

~~\\
(c) $SU(9) \to G_{SM} \times SU(3) \times U(1) \times U(1)'$

We study sfermion mass relations stemed from $[9, 6]$ of $SU(9)$ after the orbifold breaking 
$SU(9) \to G_{SM} \times SU(3) \times U(1) \times U(1)'$, which is realized by the $Z_2$ parity assignment,
\begin{eqnarray}
&~& P_0 = {\rm diag}(+1, +1, +1, +1, +1, -1, -1, -1, -1) ~, 
\label{P0-SU9} \\
&~& P_1 = {\rm diag}(+1, +1, +1, -1, -1, +1, +1, +1, -1) ~.
\label{P1-SU9}
\end{eqnarray}
After the breakdown of $SU(9)$, the sixth antisymmetric representation $[9, 6]$ with ${}_{9}C_{6}$ components
are decomposed into a sum of multiplets regarding to the subgroup $SU(3)_C \times SU(2)_L \times SU(3) \times U(1)$,
\begin{eqnarray}
[9, 6] = \sum_{l_1 =0}^{6} \sum_{l_2 = 0}^{6-l_1} \sum_{l_3 = 0}^{6-l_1-l_2} 
\left({}_{3}C_{l_1}, {}_{2}C_{l_2}, {}_{3}C_{l_3}, {}_{1}C_{6-l_1-l_2-l_3}\right) ,
\label{96}
\end{eqnarray}
where $l_1$ and $l_2$ are intergers.
The $Z_2$ parity of $\left({}_{3}C_{l_1}, {}_{2}C_{l_2}, {}_{3}C_{l_3}, {}_{1}C_{3-l_1-l_2-l_3}\right)$ is given by
\begin{eqnarray}
\mathcal{P}_0 = (-1)^{l_1+l_2} \eta_{[9,6]} ,~ \mathcal{P}_1 = (-1)^{l_1+l_3} \eta'_{[9,6]} ,
\label{Z2-SU9}
\end{eqnarray}
where $\eta_{[9,6]}$ and $\eta'_{[9,6]}$ are intrinsic $Z_2$ parities which takes a value $+1$ or $-1$.
We assume that the $Z_2$ parity (\ref{Z2-SU9}) is assigned for the left-handed Weyl fermions.
The corresponding right-handed ones have opposite $Z_2$ parities.
In Table \ref{T3}, the $Z_2$ parity ($\mathcal{P}_0, \mathcal{P}_1$) and the $U(1)'$ charge are listed
for left-handed Weyl fermions.
Each right-handed Weyl fermion has the same $U(1)'$ charge but opposite parity of corresponding left-handed one.

\begin{table}
\caption{$Z_2$ parity ($\mathcal{P}_0, \mathcal{P}_1$) and the $U(1)'$ charge for Weyl fermions from $[9,6]$}
\label{T3}
\begin{center}
\begin{tabular}{c|c|c|c|c} \hline
{\it Representation} & $\mathcal{P}_0$ & $\mathcal{P}_1$ & $U(1)$ & $U(1)'$  \\ \hline\hline
$\left({}_{3}C_{3}, {}_{2}C_{2}, {}_{3}C_1, {}_{1}C_{0}\right)_L$ & $-\eta_{[9,6]}$ & $\eta'_{[9,6]}$ & $1$ & $15$ \\ 
$\left({}_{3}C_{3}, {}_{2}C_{2}, {}_{3}C_1, {}_{1}C_{0}\right)_R$ & $\eta_{[9,6]}$ & $-\eta'_{[9,6]}$ & $1$ & $15$ \\ 
$\left({}_{3}C_{3}, {}_{2}C_{2}, {}_{3}C_0, {}_{1}C_1\right)_L$ & $-\eta_{[9,6]}$ & $-\eta'_{[9,6]}$ & $-3$ & $15$ \\
$\left({}_{3}C_{3}, {}_{2}C_{2}, {}_{3}C_0, {}_{1}C_1\right)_R$ & $\eta_{[9,6]}$ & $\eta'_{[9,6]}$ & $-3$ & $15$ \\ \hline
$\left({}_{3}C_3, {}_{2}C_{1}, {}_{3}C_2, {}_{1}C_{0}\right)_L$ & $\eta_{[9,6]}$ & $-\eta'_{[9,6]}$ & $-2$ & $6$ \\
$\left({}_{3}C_3, {}_{2}C_{1}, {}_{3}C_2, {}_{1}C_{0}\right)_R$ & $-\eta_{[9,6]}$ & $\eta'_{[9,6]}$ & $-2$ & $6$ \\
$\left({}_{3}C_3, {}_{2}C_{1}, {}_{3}C_1, {}_{1}C_1\right)_L$ & $\eta_{[9,6]}$ & $\eta'_{[9,6]}$ & $2$ & $6$ \\
$\left({}_{3}C_3, {}_{2}C_{1}, {}_{3}C_1, {}_{1}C_1\right)_R$ & $-\eta_{[9,6]}$ & $-\eta'_{[9,6]}$ & $2$ & $6$ \\
$\left({}_{3}C_{2}, {}_{2}C_2, {}_{3}C_2, {}_{1}C_{0}\right)_L$ & $\eta_{[9,6]}$ & $\eta'_{[9,6]}$ & $-2$ & $6$ \\
$\left({}_{3}C_{2}, {}_{2}C_2, {}_{3}C_2, {}_{1}C_{0}\right)_R$ & $-\eta_{[9,6]}$ & $-\eta'_{[9,6]}$ & $-2$ & $6$ \\
$\left({}_{3}C_{2}, {}_{2}C_2, {}_{3}C_1, {}_{1}C_1\right)_L$ & $\eta_{[9,6]}$ & $-\eta'_{[9,6]}$ & $2$ & $6$ \\ 
$\left({}_{3}C_{2}, {}_{2}C_2, {}_{3}C_1, {}_{1}C_1\right)_R$ & $-\eta_{[9,6]}$ & $\eta'_{[9,6]}$ & $2$ & $6$ \\ \hline
$\left({}_{3}C_3, {}_{2}C_{0}, {}_{3}C_3, {}_{1}C_{0}\right)_L$ & $-\eta_{[9,6]}$ & $\eta'_{[9,6]}$ & $3$ & $-3$ \\ 
$\left({}_{3}C_3, {}_{2}C_{0}, {}_{3}C_3, {}_{1}C_{0}\right)_R$ & $\eta_{[9,6]}$ & $-\eta'_{[9,6]}$ & $3$ & $-3$ \\ 
$\left({}_{3}C_3, {}_{2}C_{0}, {}_{3}C_{2}, {}_{1}C_1\right)_L$ & $-\eta_{[9,6]}$ & $-\eta'_{[9,6]}$ & $-1$ & $-3$ \\ 
$\left({}_{3}C_3, {}_{2}C_{0}, {}_{3}C_{2}, {}_{1}C_1\right)_R$ & $\eta_{[9,6]}$ & $\eta'_{[9,6]}$ & $-1$ & $-3$ \\ 
$\left({}_{3}C_{1}, {}_{2}C_{2}, {}_{3}C_{3}, {}_{1}C_{0}\right)_L$ & $-\eta_{[9,6]}$ & $\eta'_{[9,6]}$ & $3$ & $-3$ \\ 
$\left({}_{3}C_{1}, {}_{2}C_{2}, {}_{3}C_{3}, {}_{1}C_{0}\right)_R$ & $\eta_{[9,6]}$ & $-\eta'_{[9,6]}$ & $3$ & $-3$ \\ 
$\left({}_{3}C_{1}, {}_{2}C_{2}, {}_{3}C_{2}, {}_{1}C_{1}\right)_L$ & $-\eta_{[9,6]}$ & $-\eta'_{[9,6]}$ & $-1$ & $-3$ \\ 
$\left({}_{3}C_{1}, {}_{2}C_{2}, {}_{3}C_{2}, {}_{1}C_{1}\right)_R$ & $\eta_{[9,6]}$ & $\eta'_{[9,6]}$ & $-1$ & $-3$ \\ 
$\left({}_{3}C_{2}, {}_{2}C_{1}, {}_{3}C_{3}, {}_{1}C_{0}\right)_L$ & $-\eta_{[9,6]}$ & $-\eta'_{[9,6]}$ & $3$ & $-3$ \\ 
$\left({}_{3}C_{2}, {}_{2}C_{1}, {}_{3}C_{3}, {}_{1}C_{0}\right)_R$ & $\eta_{[9,6]}$ & $\eta'_{[9,6]}$ & $3$ & $-3$ \\ 
$\left({}_{3}C_{2}, {}_{2}C_{1}, {}_{3}C_{2}, {}_{1}C_{1}\right)_L$ & $-\eta_{[9,6]}$ & $\eta'_{[9,6]}$ & $-1$ & $-3$ \\ 
$\left({}_{3}C_{2}, {}_{2}C_{1}, {}_{3}C_{2}, {}_{1}C_{1}\right)_R$ & $\eta_{[9,6]}$ & $-\eta'_{[9,6]}$ & $-1$ & $-3$ \\ \hline
$\left({}_{3}C_{2}, {}_{2}C_{0}, {}_{3}C_{3}, {}_{1}C_{1}\right)_L$ & $\eta_{[9,6]}$ & $-\eta'_{[9,6]}$ & $0$ & $-12$ \\ 
$\left({}_{3}C_{2}, {}_{2}C_{0}, {}_{3}C_{3}, {}_{1}C_{1}\right)_R$ & $-\eta_{[9,6]}$ & $\eta'_{[9,6]}$ & $0$ & $-12$ \\ 
$\left({}_{3}C_{0}, {}_{2}C_{2}, {}_{3}C_{3}, {}_{1}C_{1}\right)_L$ & $\eta_{[9,6]}$ & $-\eta'_{[9,6]}$ & $0$ & $-12$ \\ 
$\left({}_{3}C_{0}, {}_{2}C_{2}, {}_{3}C_{3}, {}_{1}C_{1}\right)_R$ & $-\eta_{[9,6]}$ & $\eta'_{[9,6]}$ & $0$ & $-12$ \\ 
$\left({}_{3}C_{1}, {}_{2}C_{1}, {}_{3}C_{3}, {}_{1}C_{1}\right)_L$ & $\eta_{[9,6]}$ & $\eta'_{[9,6]}$ & $0$ & $-12$ \\ 
$\left({}_{3}C_{1}, {}_{2}C_{1}, {}_{3}C_{3}, {}_{1}C_{1}\right)_R$ & $-\eta_{[9,6]}$ & $-\eta'_{[9,6]}$ & $0$ & $-12$ \\ \hline
\end{tabular}
\end{center}
\end{table}

Let us take $\eta_{[9,6]} = 1$ and $\eta'_{[9,6]} =1$.
In this case, sfermions with even $Z_2$ parities are listed in Table \ref{T4}.

\begin{table}
\caption{Sfermions with even $Z_2$ parity from $[9, 6]$}
\label{T4}
\begin{center}
\begin{tabular}{c|c|c|c} \hline
{\it Representation} & {\it Species} & $U(1)$ & $U(1)'$ \\ \hline
$\left({\bf{1}}, {\bf{2}}, {\bf{3}}, {\bf{1}}\right)$ & $\tilde{l}_{1}$, $\tilde{l}_{2}$, $\tilde{l}_{3}$ & $-2$ & 6 \\ 
$\left(\overline{\bf{3}}, {\bf{1}}, \overline{\bf{3}}, {\bf{1}}\right)$ & $\tilde{d}^*_R$, $\tilde{s}^*_R$, $\tilde{b}^*_R$ & $2$ & 6 \\
$\left({\bf{1}}, {\bf{1}}, {\bf{3}}, {\bf{1}}\right)$ & $\tilde{e}^*_R$, $\tilde{\mu}^*_R$, $\tilde{\tau}^*_R$ & $1$ & 3 \\
$\left(\overline{\bf{3}}, {\bf{1}}, {\bf{3}}, {\bf{1}}\right)$ & $\tilde{u}^*_R$, $\tilde{c}^*_R$, $\tilde{t}^*_R$ & $1$ & 3 \\
$\left({\bf{3}}, {\bf{2}}, {\bf{1}}, {\bf{1}}\right)$ & $\tilde{q}_{1}$ (or $\tilde{q}_{2}$, $\tilde{q}_{3}$) & $-3$ & 3 \\
$\left({\bf{3}}, {\bf{2}}, {\bf{1}}, {\bf{1}}\right)$ & $\tilde{q}_{2}$ (or $\tilde{q}_{1}$, $\tilde{q}_{3}$) & $0$ & $-12$ \\ \hline
\end{tabular}
\end{center}
\end{table}

After the breakdown of $SU(3) \times U(1) \times U(1)'$ gauge symmetry, 
there exist the following relations at a breaking scale $M_U$,
\begin{eqnarray}
&~& {m}_{\tilde{l}_{1}}^{2}(M_U) = m_{[9,6]}^{2} + D_1 + D_2 - 2 D + 6 D' ,
\label{SU(8)-l1}\\
&~& {m}_{\tilde{l}_{2}}^{2}(M_U) = m_{[9,6]}^{2} - D_1 + D_2 - 2 D + 6 D' ,
\label{SU(8)-l2}\\
&~& {m}_{\tilde{l}_{3}}^{2}(M_U) = m_{[9,6]}^{2} - 2 D_2 - 2 D + 6 D' ,
\label{SU(8)-l3}\\
&~& {m}_{\tilde{d}^*_R}^{2}(M_U) = m_{[9,6]}^{2} - D_1 - D_2 + 2 D + 6 D' ,
\label{SU(8)-d}\\
&~& {m}_{\tilde{s}^*_R}^{2}(M_U) = m_{[9,6]}^{2} + D_1 - D_2 + 2 D + 6 D' ,
\label{SU(8)-s}\\
&~& {m}_{\tilde{b}^*_R}^{2}(M_U) = m_{[9,6]}^{2} + 2 D_2 + 2 D + 6 D' ,
\label{SU(8)-b}\\
&~& {m}_{\tilde{e}^*_{R}}^{2}(M_U) = {m}_{\tilde{u}^*_{R}}^{2}(M_U) = m_{[9,6]}^{2} + D_1 + D_2 + D + 3 D' ,
\label{SU(8)-u}\\
&~& {m}_{\tilde{\mu}^*_{R}}^{2}(M_U) = {m}_{\tilde{c}^*_{R}}^{2}(M_U) = m_{[9,6]}^{2} - D_1 + D_2 + D + 3 D' ,
\label{SU(8)-c}\\
&~& {m}_{\tilde{\tau}^*_{R}}^{2}(M_U) = {m}_{\tilde{t}^*_{R}}^{2}(M_U) = m_{[9,6]}^{2} - 2 D_2 + D + 3 D' ,
\label{SU(8)-t}\\
&~& {m}_{\tilde{q}_{1}}^{2}(M_U) = m_{[9,6]}^{2} - 3 D + 3 D' ,
\label{SU(8)-qi}\\
&~& {m}_{\tilde{q}_{2}}^{2}(M_U) = m_{[9,6]}^{2}  - 12 D' ,
\label{SU(8)-qi'}
\end{eqnarray}
where $m_{[9,6]}$ is a soft SUSY breaking scalar mass parameter, 
$D_1$, $D_2$ and $D$ are parameters which represent $D$-term condensations related to $SU(3) \times U(1)$ generator
and $D'$ stands for the $D$-term contribution of $U(1)'$.
By eliminating those five unknown parameters, we obtain the following relations at $M_U$,
\begin{eqnarray}
\hspace{-0mm}&~& {m}_{\tilde{u}^*_{R}}^{2}(M_U) = {m}_{\tilde{e}^*_{R}}^{2}(M_U) ,~ 
{m}_{\tilde{c}^*_{R}}^{2}(M_U) = {m}_{\tilde{\mu}^*_{R}}^{2}(M_U) ,~
\nonumber \\
\hspace{-0mm}&~& {m}_{\tilde{t}^*_{R}}^{2}(M_U) = {m}_{\tilde{\tau}^*_{R}}^{2}(M_U) ,
\label{SU(9)-0}\\
\hspace{-0mm}&~& {m}_{\tilde{d}^*_R}^{2}(M_U) + {m}_{\tilde{l}_{1}}^{2}(M_U) = {m}_{\tilde{s}^*_R}^{2}(M_U) + {m}_{\tilde{l}_{2}}^{2}(M_U) 
\nonumber \\
\hspace{-0mm}&~& ~~~~~~~~~~~~~~~~~~~~~~~~~~~~ = {m}_{\tilde{b}^*_R}^{2}(M_U) + {m}_{\tilde{l}_{3}}^{2}(M_U)  ,
\label{SU(9)-1}\\
\hspace{-0mm}&~& {m}_{\tilde{d}^*_R}^{2}(M_U) + {m}_{\tilde{u}^*_{R}}^{2}(M_U) = {m}_{\tilde{s}^*_R}^{2}(M_U) + {m}_{\tilde{c}^*_{R}}^{2}(M_U) 
\nonumber \\
\hspace{-0mm}&~& ~~~~~~~~~~~~~~~~~~~~~~~~~~~~~ = {m}_{\tilde{b}^*_R}^{2}(M_U) + {m}_{\tilde{t}^*_{R}}^{2}(M_U)  ,
\label{SU(9)-2}\\
\hspace{-0mm}&~& \sum_{i=1}^3 {m}_{\tilde{l}_{i}}^{2}(M_U) + {m}_{\tilde{u}^*_{R}}^{2}(M_U) 
 + {m}_{\tilde{c}^*_{R}}^{2}(M_U) + {m}_{\tilde{t}^*_{R}}^{2}(M_U) 
\nonumber \\
\hspace{-0mm}&~& ~~~~~~~ = 3 {m}_{\tilde{q}_{1}}^{2}(M_U) + {m}_{\tilde{d}^*_{R}}^{2}(M_U) 
 + {m}_{\tilde{s}^*_{R}}^{2}(M_U) + {m}_{\tilde{b}^*_{R}}^{2}(M_U) ,
\label{SU(9)-3}\\
\hspace{-0mm}&~& \sum_{i=1}^3 {m}_{\tilde{l}_{i}}^{2}(M_U) 
 + 4\left({m}_{\tilde{d}^*_{R}}^{2}(M_U) + {m}_{\tilde{s}^*_{R}}^{2}(M_U) + {m}_{\tilde{b}^*_{R}}^{2}(M_U)\right)
\nonumber \\
\hspace{-0mm}&~& ~~~~~ + 3 {m}_{\tilde{q}_{2}}^{2}(M_U) = 6\left({m}_{\tilde{u}^*_{R}}^{2}(M_U) + {m}_{\tilde{c}^*_{R}}^{2}(M_U) 
 + {m}_{\tilde{t}^*_{R}}^{2}(M_U)\right) .~~~
\label{SU(9)-4}
\end{eqnarray}
The relation (\ref{SU(9)-0}) leads to the type IIB sfermion sum rules,
\begin{eqnarray}
\hspace{-10mm} &~& M_{\tilde{u}_{R}}^{2} - M_{\tilde{e}_{R}}^{2} = M_{\tilde{c}_{R}}^{2} - M_{\tilde{\mu}_{R}}^{2}  
\nonumber \\
\hspace{-10mm} &~& ~~~ = M_{\tilde{t}_{R}}^{2} - M_{\tilde{\tau}_{R}}^{2} - m_t^2 + 2 F_t -2F_{\tau}
\nonumber \\
\hspace{-10mm} &~& ~~~ = \left(\zeta_3-20\zeta_1\left(\frac{\alpha_1}{\alpha_3}\right)\right)M_{\tilde{g}}^2
+\left(-\frac{5}{3}M_{W}^{2}+\frac{5}{3}M_{Z}^{2}\right)\cos 2\beta -10 \mathcal{S} .
\label{SU9-0} 
\end{eqnarray}
The other relations lead to following six sum rules,
\begin{eqnarray}
\hspace{-10mm} &~& M_{\tilde{d}_{R}}^{2} + M_{\tilde{e}_{L}}^{2} =  M_{\tilde{s}_{R}}^{2} + M_{\tilde{\mu}_{L}}^{2} 
 = M_{\tilde{b}_{R}}^{2} + M_{\tilde{\tau}_{L}}^{2} + 2 F_b + F_{\tau} ,
\label{SU9-1}\\
\hspace{-10mm} &~& M_{\tilde{d}_{R}}^{2} + M_{\tilde{u}_{R}}^{2} =  M_{\tilde{s}_{R}}^{2} + M_{\tilde{c}_{R}}^{2} 
 = M_{\tilde{b}_{R}}^{2} + M_{\tilde{t}_{R}}^{2} - m_t^2 + 2 F_t + 2 F_b ,
\label{SU9-2}\\
\hspace{-10mm} &~& M_{\tilde{e}_{L}}^{2} + M_{\tilde{\mu}_{L}}^{2} + M_{\tilde{\tau}_{L}}^{2} 
+ M_{\tilde{u}_{R}}^{2} + M_{\tilde{c}_{R}}^{2} + M_{\tilde{t}_{R}}^{2} 
\nonumber \\
\hspace{-10mm} &~& ~~~~ = 3M_{\tilde{u}_{L}}^{2} + M_{\tilde{d}_{R}}^{2} + M_{\tilde{s}_{R}}^{2} + M_{\tilde{b}_{R}}^{2} 
 - 3 \left(\zeta_3-20\zeta_1\left(\frac{\alpha_1}{\alpha_3}\right)\right)M_{\tilde{g}}^2
\nonumber \\ 
\hspace{-10mm} &~& ~~~~~~~~  - \left(8M_W^2 - 5M_Z^2\right)\cos 2\beta + m_t^2 - 2 F_t + 2 F_b - F_{\tau} ,
\label{SU9-3}\\
\hspace{-10mm} &~& M_{\tilde{e}_{L}}^{2} + M_{\tilde{\mu}_{L}}^{2} + M_{\tilde{\tau}_{L}}^{2} 
 + 4\left(M_{\tilde{d}_{R}}^{2} + M_{\tilde{s}_{R}}^{2} + M_{\tilde{b}_{R}}^{2}\right)
 + 3M_{\tilde{c}_{L}}^{2} 
\nonumber \\
\hspace{-10mm} &~& ~~~~ = 6\left(M_{\tilde{u}_{R}}^{2} + M_{\tilde{c}_{R}}^{2} + M_{\tilde{t}_{R}}^{2}\right) 
 - 3 \left(\zeta_3 - 2 \zeta_2 \left(\frac{\alpha_2}{\alpha_3}\right)^2 + 70 \zeta_1 
 \left(\frac{\alpha_1}{\alpha_3}\right)^2\right) M_{\tilde{g}}^2
\nonumber \\ 
\hspace{-10mm} &~& ~~~~~~~~  - 9 \left(M_W^2 - M_Z^2\right)\cos 2\beta - 6 m_t^2 + 12 F_t- 8 F_b - F_{\tau} .
\label{SU9-4}
\end{eqnarray}
The brane fields live on the orbifold fix points and respect with the gauge symmetry on the brane.
Therefore a part of type IIA sum rules hold on for brane fields if our world is a boundary with $y = 0$.

\section{Conclusions}

We have studied the sparticle spectrum in the MSSM and derived sum rules among sparticle masses
under the assumption that models beyond the MSSM are four-dimensional SUSY GUTs 
or five-dimensional SUSY orbifold GUTs.
We have obtained various kinds of scalar sum rules and we can classify them into following four types.

~~\\
(Type I) Sfermion sum rules inherent the MSSM:
\begin{eqnarray}
&~& M_{\tilde{u}_L}^2 - M_{\tilde{d}_L}^2 = M_{\tilde{\nu}_{eL}}^2 - M_{\tilde{e}_L}^2 = M_{\tilde{c}_L}^2 - M_{\tilde{s}_L}^2 
= M_{\tilde{\nu}_{\mu L}}^2 - M_{\tilde{\mu}_L}^2 
\nonumber \\
&~& ~~~ = M_{\tilde{t}_L}^2 - M_{\tilde{b}_L}^2 - m_t^2 =
 M_{\tilde{\nu}_{\tau L}}^2 - M_{\tilde{\tau}_L}^2 = M_W^2 \cos2\beta .
\label{I} 
\end{eqnarray}
These sum rules are derived from the fact that left-handed fermions (and its superpartner) form $SU(2)_L$ doublets,
and they are irrelevant to the structure of models beyond the MSSM.
The sfermion sector (and the breakdown of electroweak symmetry) in the MSSM can be tested by using them.

~~\\
(Type IIA) Sfermion sum rules from four-dimensional grand unification:
\begin{eqnarray}
\hspace{-5mm}&~& M_{\tilde{u}_{L}}^{2}-M_{\tilde{u}_{R}}^{2} = M_{\tilde{c}_{L}}^{2}-M_{\tilde{c}_{R}}^{2} 
= M_{\tilde{t}_{L}}^{2}-M_{\tilde{t}_{R}}^{2} - F_t + F_b 
\nonumber \\
\hspace{-5mm}&~& ~~~ = \left(\zeta_2 \left(\frac{\alpha_2}{\alpha_3}\right)^2 - 15\zeta_1 \left(\frac{\alpha_1}{\alpha_3}\right)^2\right)
M_{\tilde{g}}^2
\nonumber \\
\hspace{-5mm}&~& ~~~~~~~~~~~~~~~~~~ +\left(\frac{4}{3}M_{W}^{2}-\frac{5}{6}M_{Z}^{2}\right)\cos 2\beta +5 \mathcal{S} , 
\label{IIA-1} \\
\hspace{-5mm}&~& M_{\tilde{u}_{R}}^{2} - M_{\tilde{e}_{R}}^{2} = M_{\tilde{c}_{R}}^{2} - M_{\tilde{\mu}_{R}}^{2}  
= M_{\tilde{t}_{R}}^{2} - M_{\tilde{\tau}_{R}}^{2} - m_t^2 + 2 F_t -2F_{\tau}
\nonumber \\
&~& ~~~ = \left(\zeta_3 -20\zeta_1\left(\frac{\alpha_1}{\alpha_3}\right)^2\right)
 M_{\tilde{g}}^2 
\nonumber \\
\hspace{-5mm}&~& ~~~~~~~~~~~~~~~~~~ +\left(-\frac{5}{3}M_{W}^{2}+\frac{5}{3}M_{Z}^{2}\right)\cos 2\beta -10 \mathcal{S} ,
\label{IIA-2} \\
\hspace{-5mm}&~& M_{\tilde{e}_{L}}^{2}-M_{\tilde{d}_{R}}^{2} = M_{\tilde{\mu}_{L}}^{2}-M_{\tilde{s}_{R}}^{2} 
= M_{\tilde{\tau}_{L}}^{2}-M_{\tilde{b}_{R}}^{2} + F_{\tau} - 2 F_b
\nonumber \\
\hspace{-5mm}&~& ~~~ = \left(- \zeta_3 + \zeta_2 \left(\frac{\alpha_2}{\alpha_3}\right)^2 + 5\zeta_1 \left(\frac{\alpha_1}{\alpha_3}\right)^2\right)
M_{\tilde{g}}^2 
\nonumber \\
\hspace{-5mm}&~& ~~~~~~~~~~~~~~~~~~ +\left(-\frac{4}{3}M_{W}^{2}+\frac{5}{6}M_Z^2\right)\cos 2\beta - 5 \mathcal{S} .
\label{IIA-3}
\end{eqnarray}
These sum rules are derived in the case with the direct breakdown of
a grand unified symmetry to the SM one,
and hence the sfermion sector (and the grand unification) in SUSY GUT can be tested by using them.

~~\\
(Type IIB) Sfermion sum rules from five-dimensional orbifold grand unification:
\begin{eqnarray}
 M_{\tilde{u}_{R}}^{2} - M_{\tilde{e}_{R}}^{2} 
 = \left(\zeta_3 -20\zeta_1\left(\frac{\alpha_1}{\alpha_3}\right)^2\right) M_{\tilde{g}}^2 
  -\frac{5}{3}\left(M_{W}^{2} - M_{Z}^{2}\right)\cos 2\beta -10 \mathcal{S} .
\label{IIB} 
\end{eqnarray}
This sum rule is a piece of type IIA ones and
it is derived on the orbifold breaking of $SU(N)$ gauge symmetry for bulk fields with 
an antisymmetric representation if the bulk field contains ${\bf 10}_L$ or ${\overline{\bf 10}}_R$ under the subgroup $SU(5)$,
and $SU(2)_L$ singlets have even $Z_2$ parities.
The orbifold breaking (of $SU(N)$ gauge symmetry) can be tested by using it.

~~\\
(Type III) Sfermion sum rules specific to each model:\\
There are sfermion sum rules specific to physics beyond the MSSM
such as the chain breaking of gauge symmetry or representations of gauge group in SUSY orbifold GUT.
We refer them as type III sfermion sum rules.

~~\\
(Type IV) Sfermion sum rules favored by phenomenology:\\
There are sfermion sum rules favored from the viewponit of particle phenomenology.
For example, if sum rules such as $M_{\tilde{u}_{L}}^{2} = M_{\tilde{c}_{L}}^{2}$, 
$M_{\tilde{d}_{L}}^{2} = M_{\tilde{s}_{L}}^{2}$ and so on hold on, the sufficient suppression of FCNC processes
can be realized.
We refer them as type IV sfermion sum rules.\\

Four-dimensional superstring models, in general, possess extra gauge symmetries including several $U(1)$s
and sfermions could have a different quantum numbers.
Therefore sum rules different from those in our analysis could be derived.
In fact, some exotic scalar mass relations have been obtained.\cite{KK&K}
Superstring theory can be tested by using them.
In any case, sparticle sum rules can play as useful probes of the MSSM and beyond.

Some assumptions in our analysis might be too strong to describe a high-energy physics correctly.
The resultant sum rules can be modified by relaxing some assumptions.
For example, if extra gauge symmetries are broken at different scales from the grand unification scale or the orbifold
breaking scale, or if extra particles exist in the intermediate scale,
soft SUSY breaking parameters receive extra renormalization effects and then our analysis should be modified.
In the case that effects such as threshold corrections, 
higher loop RG evolution, $F$-term contributions and/or higher dimensional operators are sizable, 
we should consider them.

In our analysis, we have assumed the gravity-mediated SUSY breaking in the case that the dynamics in the hidden sector 
do not give sizable effects.
It is also important to study sum rules specific to the way of mediation and the hidden dynamics. 
Sum rules have been examined in the gauge-mediated SUSY breaking models\cite{gauge-mediation},
the anomaly-mediated SUSY breaking models\cite{anomaly-mediation} 
and models with strong dynamics in the hidden sector.\cite{CR&S}

In most cases, we have not required the mass degeneracy for each squark and slepton species in the first two generations
and then dangerous FCNC processes can be induced unless those masses are rather heavy or fermion
and its superpartner mass matrices are aligned.
In orbifold family unification models, a non-ableian subgroup such as $SU(3)$ of $SU(N)$ plays the role of family symmetry
and its $D$-term contributions spoil the mass degeneracy.
Conversely, the requirement of degenerate masses would give a constraint on the $D$-term condensations
and lead to extra sfermion sum rules.
It is important to study phenomenilogical aspects of orbifold GUTs
and construct a realistic model which explains family structure.
It is interesting to derive sfermion sum rules specific to each model and classify them. 
These will be presented in a separate publication.

\section*{Acknowledgements}
This work was supported in part by Scientific Grants from the Ministry of Education and Science, 
Grant No.13135217, Grant No.18204024, Grant No.18540259 (Y.K.).

\end{document}